\documentclass[a4paper, final,leqno,onefignum,onetabnum]{siamltex1213C}

\usepackage{graphicx}
\usepackage{caption}
\usepackage{subcaption}
\usepackage{amsfonts}
\usepackage{amsmath}
\usepackage{amssymb}
\usepackage[dvips]{graphicx}
\usepackage{dcolumn}
\usepackage{dsfont}
\usepackage{times}
\usepackage{epsfig}
\usepackage{xcolor}

\def\bxi {{\boldsymbol\xi}}
\def\bzeta {{\boldsymbol\zeta}}
\def\bkappa {{\boldsymbol\kappa}}

\def\bx {{\bf x}}

\def\bX {{\bf X}}

\newcommand{\beq}{\begin{equation}}
\newcommand{\eeq}{\end{equation}}
\newcommand{\ba}{\begin{eqnarray}}
\newcommand{\ea}{\end{eqnarray}}
\newcommand{\cosec}{\operatorname{cosec}}

\usepackage{caption}

\begin{document}
\title{Wave mechanics in media pinned at Bravais lattice points\thanks{This work was
supported by the EPSRC and ERC}}


\author{M. Makwana\footnotemark[2] \and T. Antonakakis\footnotemark[2]
  \and B. Maling\footnotemark[2] \and S. Guenneau\footnotemark[3] \and R.~V. Craster\footnotemark[2]}
\maketitle

\renewcommand{\thefootnote}{\fnsymbol{footnote}} \footnotetext[2]{Department of Mathematics, Imperial College London, London SW7 2AZ, UK}
\footnotetext[3]{Institut Fresnel, UMR CNRS 7249, University of Aix Marseille, Marseille, France} 
\renewcommand{\thefootnote}{\arabic{footnote}}

\begin{abstract}
 
The propagation of waves through
microstructured media with
periodically arranged inclusions has applications in many areas of physics and engineering, 
stretching from photonic crystals through
to seismic metamaterials. In the high-frequency regime, modelling such behaviour is complicated by multiple scattering of
 the resulting short waves between the inclusions.
Our aim is to develop an asymptotic
theory for modelling systems with arbitrarily-shaped inclusions located on general Bravais lattices. We then consider the limit of point-like inclusions, the advantage being that exact solutions can be obtained using Fourier methods, and go on to derive effective medium equations using asymptotic analysis. This approach allows us to explore the
underlying reasons for dynamic anisotropy, localisation of
waves, and other properties typical of such systems, and in particular their dependence
upon geometry.  
Solutions of the effective medium equations are compared with the exact solutions,
 shedding further light on the underlying physics. We focus on examples that
 exhibit dynamic anisotropy as these demonstrate the capability of the asymptotic theory
 to pick up detailed qualitative and quantitative features.  
 
\end{abstract}

\begin{keywords}
Homogenisation, Bloch waves, Multiple-scales
\end{keywords}

\section{Introduction}

The design of  structures that have material properties that
 are controllable, or that do not naturally occur, is highly
 topical. It is now possible to talk of a negative refractive
 index, and to design materials accordingly in acoustics \cite{craster12a} and
 electromagnetism, or to design photonic crystals \cite{joannopoulos08a} whose overall properties are
 governed by their regular microstructure. Much of this literature draws
 upon earlier work in solid state physics \cite{brillouin53a, kittel96a}.

\begin{figure}
\centering
\begin{subfigure}[c]{0.41\linewidth}
\caption{\small {
Oblique}}
\includegraphics[width=\textwidth]{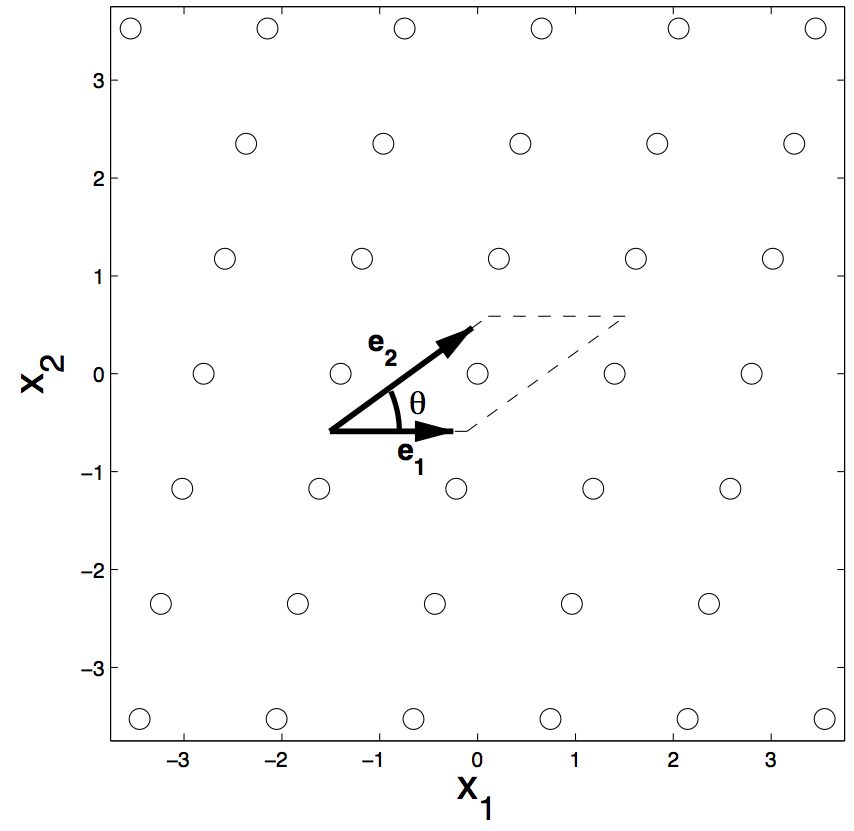}
\end{subfigure}
\begin{subfigure}[c]{0.42\linewidth}
\caption{\small {Rhombic}}
\includegraphics[width=\textwidth]{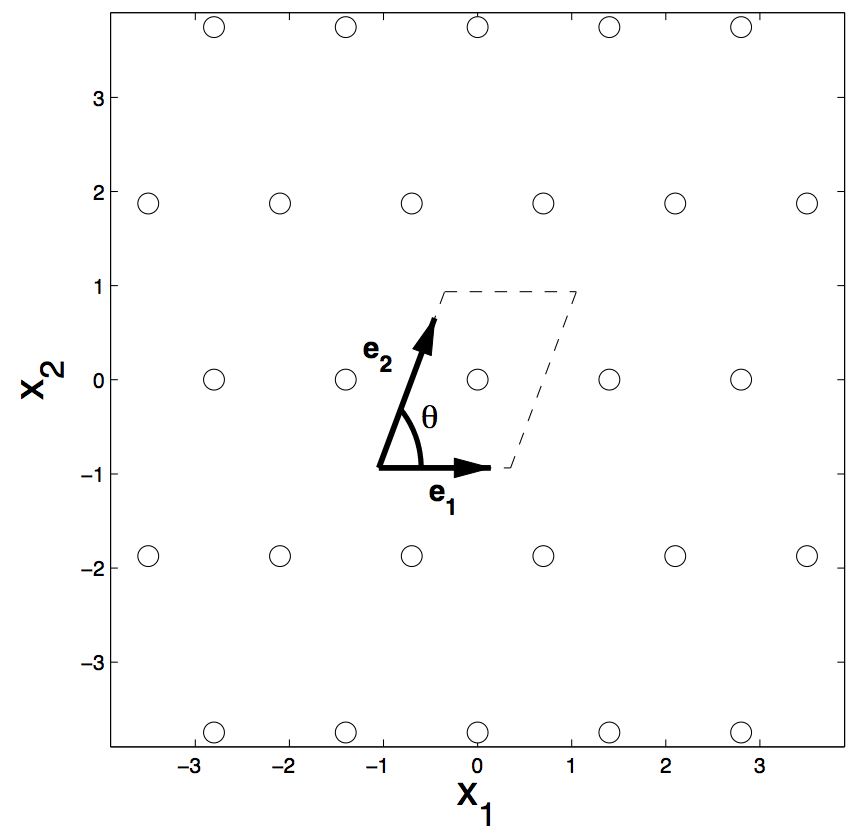}
\end{subfigure}
\qquad
\vspace{0.5cm}
\begin{subfigure}[c]{0.41\linewidth}
\caption{\small {
Rectangular}}
\includegraphics[width=\textwidth]{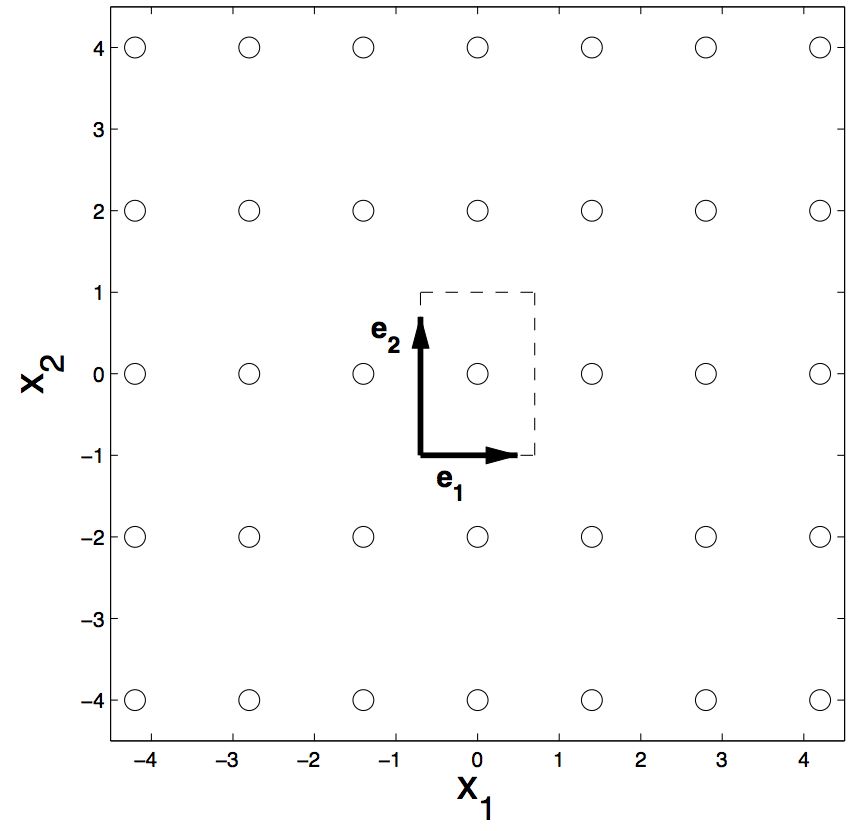}
\end{subfigure}
\begin{subfigure}[c]{0.41\linewidth}
\caption{\small {
Square}}
\includegraphics[width=\textwidth]{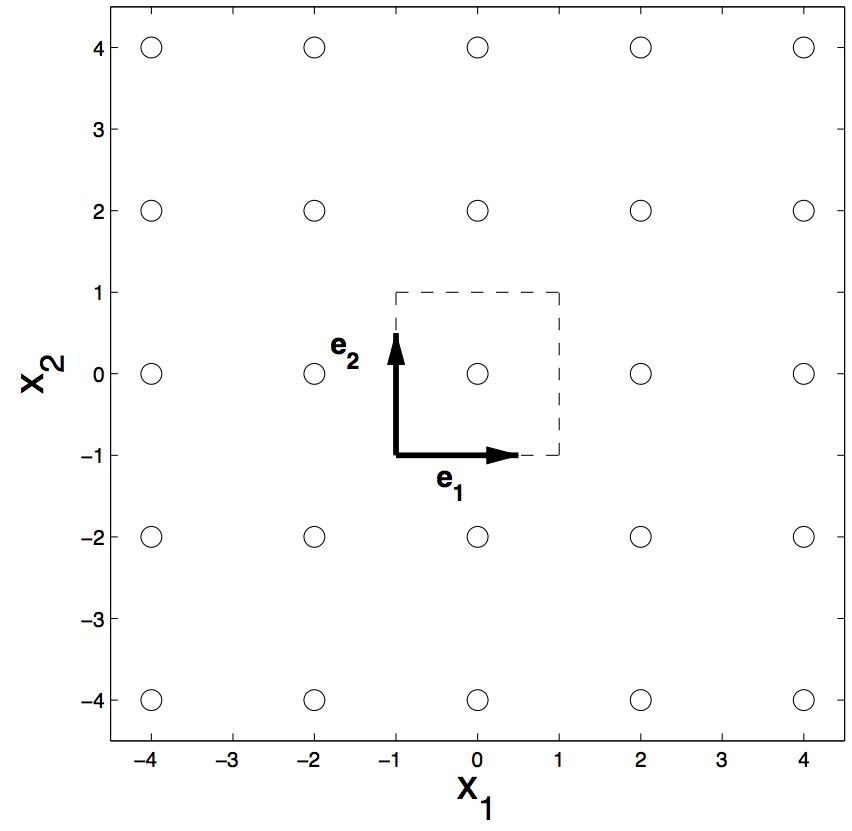}
\end{subfigure}
\qquad
\vspace{0.5cm}
\begin{subfigure}[c]{0.41\linewidth}
\caption{\small {
Hexagon}}
\includegraphics[width=\textwidth]{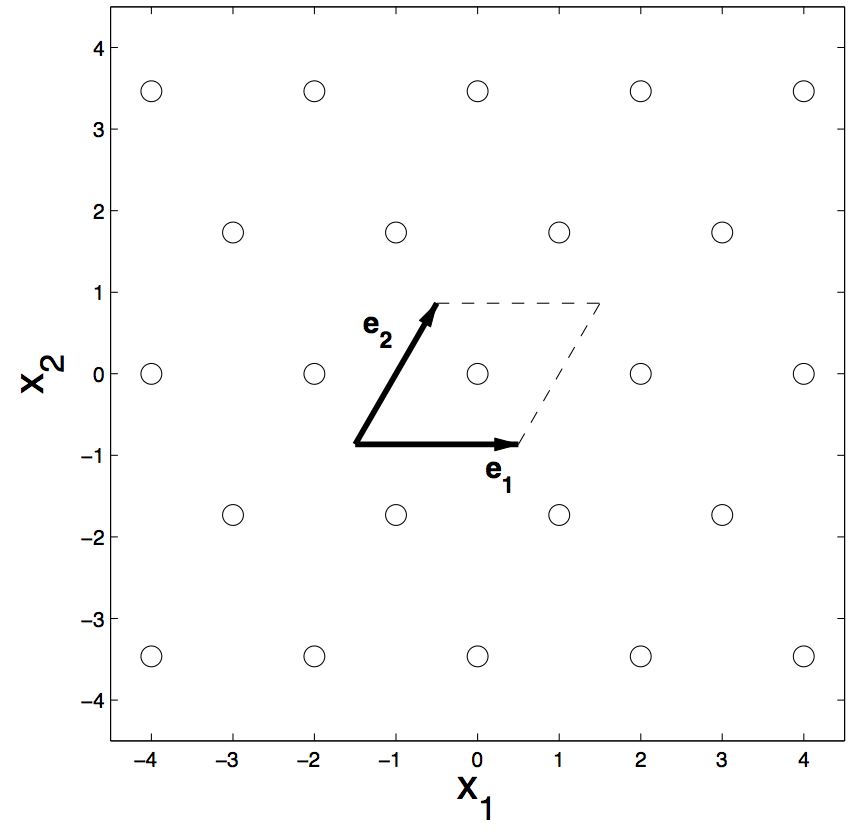}
\end{subfigure}
\caption{
\label{fig:oblique_rhombic}
\small {Illustrations of the 5 Bravais lattices together with their associated
lattice basis vectors.}}
\end{figure}

 For design purposes, the arrangement of inclusions in
 hexagonal or rhombic patterns is typical in photonic
 crystals \cite{zolla05a}. This special subset can be distinguished from the remaining oblique planar geometries by the symmetries of their first Brillouin zones (see section \ref{sec:formulation}), and together with these, and two distinct orthogonal lattices, they form the set of two-dimensional Bravais lattices \cite{brillouin53a} as shown in Fig. \ref{fig:oblique_rhombic}. Notably, a
 honeycomb array is simply a hexagonal array with two pins per elementary
 cell so we can, and do, consider this case too due to the remarkable properties
 of graphene \cite{neto09a}.
 
Our approach is to consider two distinct systems in this paper;
physically the Helmholtz equation
\beq
\nabla_{\bx}\cdot\left[\hat{a}\left(\bx \right)\nabla_{\bx} u(\bx) \right]+\hat{\rho}\left(\bx \right) \Omega^2 u(\bx)=0
\label{eq:helmholtz}
\eeq
can be used to model transverse electric (TE) or transverse magnetic (TM)
 polarised electromagnetic waves, shear horizontal (SH) 
 polarised elastic waves and pressure waves in the frequency domain. The spatially-dependent physical parameters $\hat{a}(\bx)$, $\hat{\rho}(\bx)$ will be assumed periodic in space, and represent different quantities depending on the physical setting, for example stiffness and density in the SH elastic system. 
 We consider structures containing arbitrarily shaped inclusions and provide Dirichlet conditions on their boundaries, and later take the zero radius limit to point scatterers. There is wide interest in similar continuum problems; for instance
\cite{wu12a,wu12b} provide typical applications for rhombic lattice arrangements.

A system closely related to the one described above is that of a structured elastic plate governed by the Kirchhoff-Love equation, for which vertical displacements satisfy
\beq
\nabla_{\bf x}^2\left[\beta\left(\bx \right)\nabla_{\bf x}^2 u\left(\bx \right) \right]-\mu\left(\bx \right){\hat \Omega}^2 u\left(\bx \right) = 0.
\label{eq:kirchoff}
\eeq 
The non-dimensional squared frequency is
\beq
\hat{\Omega}^2=\frac{\rho_0 A_0 l^4 \Omega^2}{E_0 I_0},
\eeq where the quantities appearing above are reference values of density, cross-sectional area, Young's modulus and cross-section moment of inertia, related to the (spatially varying) values in the plate via the relations
$EI=E_0I_0\beta\left({\bf x} \right), \rho A=\rho_0
A_0\mu\left({\bf x} \right)$. The dimensionless quantities $\beta$ and $\mu$ thus represent the non-dimensional mass per unit length and flexural rigidity of the
plate respectively. There has been considerable interest in such systems as platonic crystals, in which ideas from photonic crystals are transplanted into this setting \cite{antonakakis13b,brule14,dubois13,farhat10b,farhat10a,mcphedran15a,smith13a,torrent13a}. The additional derivatives in (\ref{eq:kirchoff}) compared to (\ref{eq:helmholtz}) lead to differences in the corresponding physical systems, but since
 the mathematics is closely related we choose to develop the theory for both equations in
 parallel. In the latter case we too shall study the zero-radius limit, noting that for an array of clamped pins there is a straightforward exact
 solution for a square array
 using Fourier series \cite{antonakakis12a,langley97b,mace96a} 
 that can be modified to other lattice arrangements
 \cite{metcalfe02a}. Since points are the zero-radius limit of circular
 holes, one can also approach this problem
 using multipole methods \cite{movchan07c}, though we do not pursue this here. Additionally the stop band and localisation effects of plates  containing an ordered arrangement of thin long fibres, which is reminiscent of clamped pins, has been studied experimentally \cite{ruppin14}.

Our primary aim is to generate homogenised, effective medium equations that
capture the essential physics in a long-scale governing equation that
encapsulates the short-scale structure in geometry-dependent
coefficients. This is achieved using the multiple-scales methodology
of high frequency homogenisation developed for square lattice structures in \cite{craster10a}, and here generalised for Bravais lattices for which there are significant technical
issues to overcome. The asymptotic methodology of \cite{craster10a}
relies on perturbing away from solutions found at the
edges of the Brillouin zone, though we demonstrate in section \ref{sec:formulation} that it can be applied at any point therein.  
The importance of band gap edges was recognised in the analysis community 
\cite{birman06a},
 as well as by those studying the high frequency long wave asymptotics of waveguides \cite{gridin05a,kaplunov89a,kaplunov05a}, a subject that has strong
 analogies 
 to wave propagation in periodic media 
\cite{craster14a}. We note that the desire to obtain effective properties is not limited to systems governed by (\ref{eq:helmholtz}), (\ref{eq:kirchoff}) and also
arises in studies of 
the Schr\"odinger equation  
\cite{allaire05a,hoefer11a}, in particular for potentials associated with honeycomb
  structures 
\cite{fefferman12a}.

An additional aim here is to generate exact Fourier series solutions in the zero-radius limit of pinned points for both systems \eqref{eq:helmholtz}, \eqref{eq:kirchoff}. We then use these solutions to investigate the effectiveness of the asymptotic method, as well as the effects of geometry, and in
particular lattice symmetries. In the case of the Kirchhoff-Love equation (\ref{eq:kirchoff}), this methodology applied to a square lattice gives effective medium equations
\cite{antonakakis12a} that predict and model shielding
and lensing effects for elastic plates \cite{antonakakis13b,antonakakis14a}. Given this success for square arrays,
  and subsequent extensions to vector wave systems such as in-plane
  elasticity \cite{antonakakis14c, boutin14a}, it is natural to extend
  the approach to Bravais lattice arrangements that are of
  broader physical interest. 


The plan of the paper is as follows: in section \ref{sec:formulation} we introduce the mathematical description of the lattice
geometries. With this background we then proceed, in section \ref{sec:asymptotic}, to our two-scale analysis of the wave systems in question. 
The upshot in both cases is that a long-scale
partial differential equation emerges with the short-scale and
geometry built into a tensor of coefficients. Notably, although the
governing equation (\ref{eq:kirchoff}) is fourth-order, 
the long-scale equation that emerges is typically of
second-order. Situations whereby dispersion curves cross are not
uncommon and often correspond  to Dirac-like points that arise due to
repeated eigenvalues; this is also incorporated into the asymptotic
theory and described herein. Section \ref{sec:exact} uses
Fourier methods to construct exact solutions in the limit of point-like inclusions, which are used to provide checks on the asymptotic theory, 
as well as being interesting in their own right. These, together with
the asymptotics, are used to interpret and model wave phenomena in section \ref{sec:results}. The asymptotic
theory clearly captures not just qualitative features, but also quantitative decay rates for localised
states and the detailed dispersive properties near standing wave
eigenfrequencies. Concluding remarks are drawn together in section
\ref{sec:conclusions}. 


\section{Formulation}
\label{sec:formulation}
We consider a two-dimensional structure comprised of a doubly periodic
array of cells. In our previous articles, inclusions were assumed to be arranged in a square geometry and our aim 
here is to create an asymptotic theory that is more general in
terms of lattices, and to emphasise the differences between orthogonal and non-orthogonal geometries. We
extend the existing methodology to deal with the $5$ fundamental
two-dimensional Bravais lattices given by the crystallographic
restriction theorem, as shown in Fig. \ref{fig:oblique_rhombic}. We allow for cases in which the
cells contain $N \in \mathbb{N}$ inclusions, which is achieved
 by associating one or more inclusions to each lattice point. 

\begin{figure} [ht!]
\centering
\begin{subfigure}[c]{0.27\linewidth}
\caption{\small {
Oblique}}
\includegraphics[width=\textwidth]{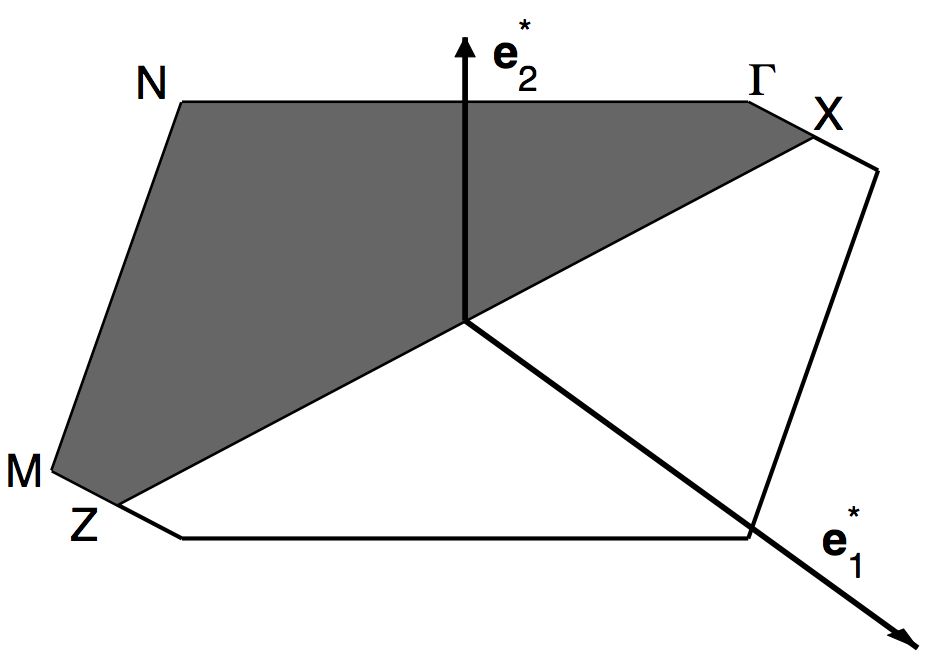}
\end{subfigure}
\begin{subfigure}[c]{0.27\linewidth}
\caption{\small {Rhombic}}
\includegraphics[width=\textwidth]{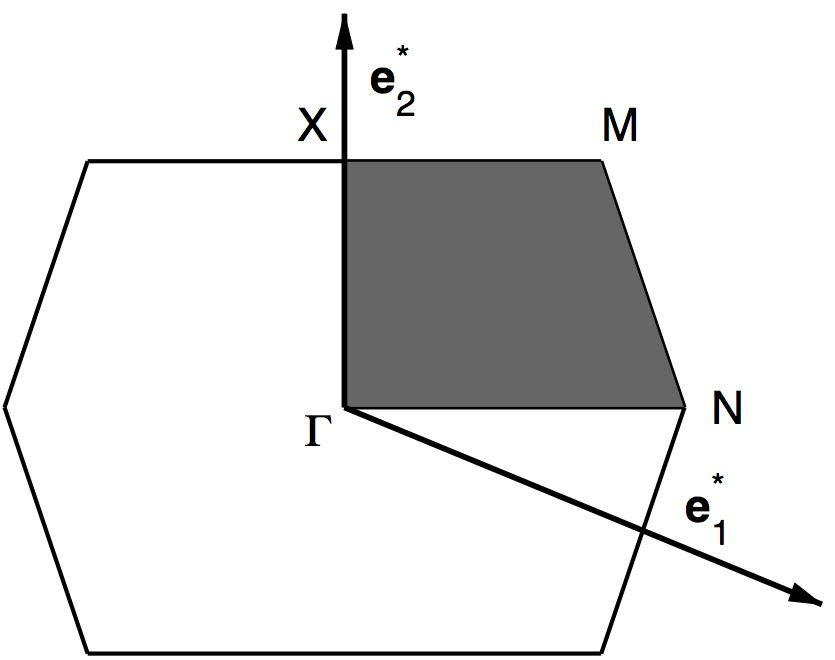}
\end{subfigure}
\qquad
\vspace{0.5cm}
\begin{subfigure}[c]{0.25\linewidth}
\caption{\small {
Rectangular}}
\includegraphics[width=\textwidth]{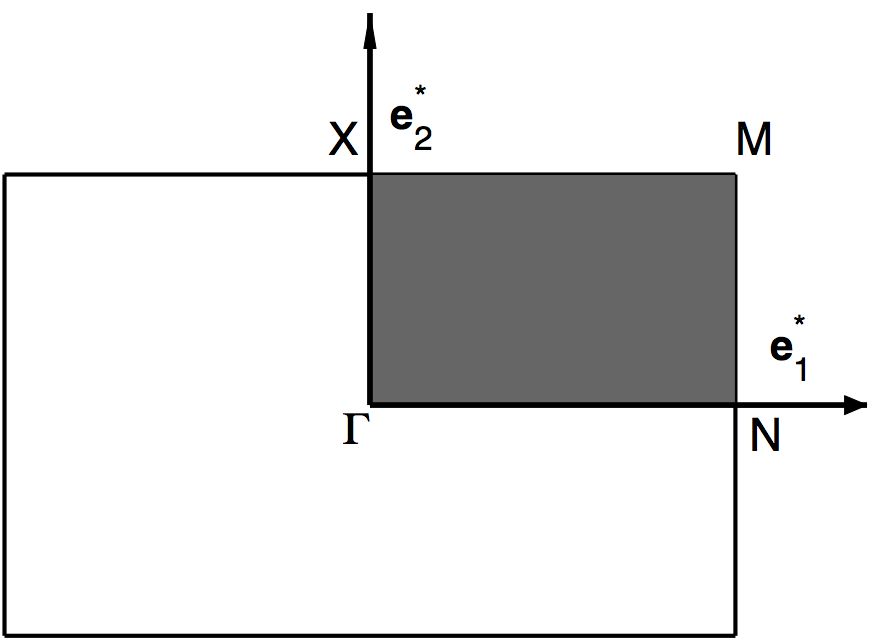}
\end{subfigure}
\\
\begin{subfigure}[c]{0.23\linewidth}
\caption{\small {
Square}}
\includegraphics[width=\textwidth]{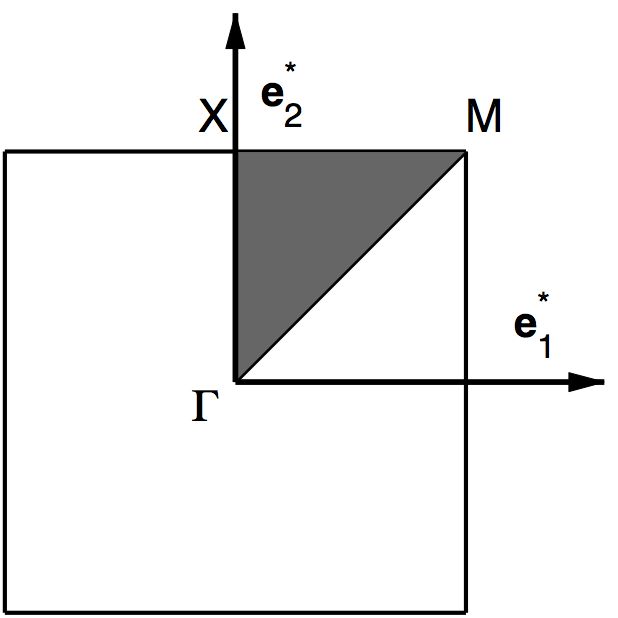}
\end{subfigure}
\qquad
\vspace{0.5cm}
\begin{subfigure}[c]{0.26\linewidth}
\caption{\small {
Hexagon}}
\includegraphics[width=\textwidth]{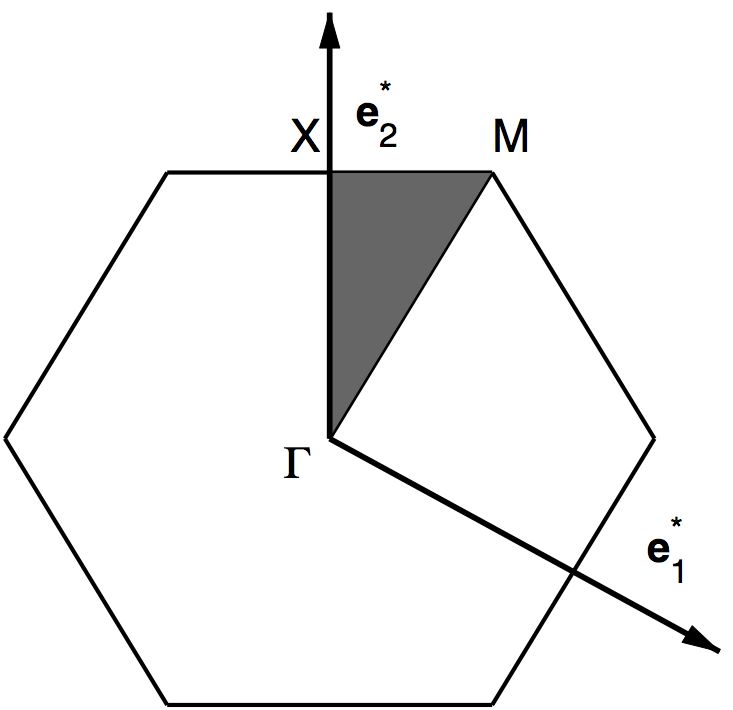}
\end{subfigure}
\caption{
\label{fig:brillouin_zones}
\small {Illustrations of the first Brillouin zones for the 5 Bravais lattices. The shaded region shows the irreducible zone. Parameter values for each of the panels are (a) $\varphi(0.7,\pi/5)$ \hspace{0.1cm} (b) $\varphi(0.7,\arccos(0.35))$ \hspace{0.1cm} (c) $\varphi(0.7,\pi/2)$ \hspace{0.1cm} (d) $\varphi(1,\pi/2)$ \hspace{0.1cm}(e) $\varphi(1,\pi/3)$.}}
\end{figure}

By definition, a Bravais lattice consists of an infinite array generated by a set of discrete translation vectors
\beq
{\bf R}=\sum\limits_{i=1}^N n_i {\bf e}_i, \hspace{0.5cm}  n_i \in \mathbb{N},
\label{eq:bravais_defn}
\eeq where ${\bf e}_i$ are the lattice basis vectors and $n_i$ are
integer weighting functions. Consequently the medium appears identical when viewed from any unit cell. Generally, the primitive vectors defining a two-dimensional periodic lattice are written as
\beq
{\bf e}_1=\alpha A_{2D} {\bf i}, \hspace{0.4cm} {\bf e}_2=A_{2D}\left[\cos\theta {\bf i}+\sin\theta {\bf j} \right],
\label{eq:gen_basis_vec} \eeq where ${\bf i},{\bf j}$ are the unit
orthogonal vectors, $\alpha$ is the asymmetry ratio, $A_{2D} \in
\mathbb{R}$, and $0 < \theta \leq \pi/2$. 
To obtain reciprocal lattice basis vectors, ${\bf e}_i^*$, we utilise the following orthogonality condition,
\beq
{\bf e}_i\cdot{\bf e}_j^*=2 \pi \delta_{ij}, \hspace{0.2cm} i,j=1,2
\eeq to give us the reciprocal lattice basis vectors
\beq
{\bf e}_1^*=\frac{2\pi}{\alpha A_{2D}}\left[ {\bf i}-\cot\theta{\bf j} \right], \hspace{0.4cm} {\bf e}_2^*=\frac{2\pi}{A_{2D}}\cosec\theta {\bf j}.
\label{eq:gen_recip_vec}
\eeq 
The non-dimensional position vector of any point in the medium is given by ${\bf r}= \xi_1 {\bf i} + \xi_2 {\bf j} = \zeta_1 {\bf e}_1 + \zeta_2 {\bf e}_2$ and using this, along with equation \eqref{eq:gen_basis_vec}, we deduce the following relations
\beq
\xi_1=A_{2D}\left[\zeta_1 \alpha + \zeta_2 \cos\theta\right], \hspace{0.4cm} \xi_2= \zeta_2 A_{2D} \sin\theta.
\label{eq:gen_coord}\eeq We denote the lattice satisfying equations
\eqref{eq:gen_basis_vec}-\eqref{eq:gen_coord} by the general function
$\varphi(\alpha,\theta)$; this definition is sufficient as any
two-dimensional lattice can be described by its  asymmetry ratio and
angular component. For example, the square lattice is denoted by
$\varphi\left(1,\pi/2 \right)$ and the remaining Bravais lattices are
defined in table \ref{bravais_def}. Clearly, lattices (b) - (e) in Fig. \ref{fig:oblique_rhombic} can be considered special cases of the oblique lattice (a), but a key distinction can be made by analysing
the symmetries of the first Brillouin zones (Fig. \ref{fig:brillouin_zones}); for oblique lattices that have parameter values $(\alpha, \theta)$
which do not coincide with those found in the $4$ distinguished lattices,
there exists only a single line of symmetry bisecting the first
Brillouin zone. This bears particular relevance to our analysis as the majority of features we are interested in, such as stop-bands, critical points and
degeneracies, are captured by traversing along the boundary of the
irreducible Brillouin zone. Notably one requires some care in doing so
\cite{craster12b}. 

\begin{table}
\begin{center}
    \begin{tabular}{ | c | c | c | p{5cm} |}
    \hline
    {\bf Topology} & {\bf General function} & {\bf Restrictions}\\ \hline
    Oblique & $\varphi\left(\alpha,\theta\right)$ & $\alpha \neq 1, \theta \neq \pi/2$  \\ \hline
    Rhombic & $\varphi\left(\alpha,\arccos\left(\alpha/2 \right)\right)$ & $\alpha \neq 1 $\\ \hline
    Rectangle & $\varphi\left(\alpha,\pi/2\right)$ & $\alpha \neq 1$ \\ \hline
    Square & $\varphi\left(1,\pi/2\right)$ & - \\ \hline
    Hexagonal & $\varphi\left(1,\pi/3\right)$ & - \\
    \hline
    \end{tabular}
    \caption{\small {The 5 fundamental Bravais lattices are defined in the above table. Definitions are given via the function $\varphi(\alpha,\theta)$, as any two-dimensional lattice can be uniquely defined by the angular component $\theta$ and asymmetry ratio $\alpha$.}}
        \label{bravais_def}
        \end{center}
\end{table}

The boundary of the first Brillouin zone, for all the geometries, is given by 
\beq
{\bf G}\cdot\left({\bkappa}-\frac{{\bf G}}{2} \right)=0, \hspace{0.2cm} {\bf G}=m_1{\bf e}_1^*+m_2{\bf e}_2^*,
\label{eq:perp_bi_defn} \eeq 
where $m_i=(0,\pm 1)$. This definition allows us to find a general form for the edges of the irreducible zone, $\Gamma, X,M$ and $N$, for the rhombic, hexagonal and orthogonal geometries (fig. \ref{fig:brillouin_zones}):
\begin{gather}
\Gamma=\left(0,0 \right), \enskip X=\left(0,\pi\cosec(\theta) \right),\nonumber\\
M=\left(\frac{\pi}{\alpha}\left[1+\cot(\theta)^2 \right]-\pi\cot(\theta)\cosec(\theta),\pi\cosec(\theta) \right), \nonumber\\
N=\left(\frac{\pi}{\alpha}\left[1-\cot(\theta)^2 \right]+\pi\cot(\theta)\cosec(\theta),\pi \left[\cosec(\theta)-\frac{2}{\alpha}\cot(\theta)\right] \right).
\label{eq:gen_bdry_pts}
\end{gather} 
Given this geometrical set-up, and the relation to the reciprocal
space, 
we now move on to the asymptotic procedure. 

\section{Asymptotic theory}
\label{sec:asymptotic}
In the following section we detail the generalised high frequency homogenisation method. Initially we consider the Helmholtz equation, before moving onto the case of platonic crystals governed by the Kirchhoff-Love equation.

\subsection{Helmholtz equation}
In this section we detail the asymptotic procedure as applied to the following governing equation:
\beq
\nabla_{\bx}\cdot\left[\hat{a}\left(\bx \right)\nabla_{\bx} u(\bx) \right]+\Omega^2\hat{\rho}\left(\bx \right)u(\bx)=0.
\label{eq:gov_eqn}\eeq 
Here ${\bx}$ are the orthogonal coordinates on the infinite domain. The material is characterised by the periodic functions $\hat{a}\left(\bx \right),\hat{\rho}\left(\bx \right)$.

In the dimensional setting, the unit cell is taken to have sides of
length $2\alpha l$ and $2 l$ in the ${\bf e}_1, {\bf e}_2$ directions
respectively, where for convenience we also set $A_{2D}=1$ in
equations \eqref{eq:gen_basis_vec}. We begin to non-dimensionalise equation \eqref{eq:gov_eqn}
by setting, $\hat{a} \equiv\hat{a}_{0}a(\bx)$ and
$\hat{\rho}\equiv\hat{\rho}_{0}\rho(\bx)$ in (\ref{eq:gov_eqn}):
\beq
l^2\nabla_{\bx}\cdot
[a(\bxi)\nabla_{\bx}u(\bx)]+\Omega^2\rho(\bxi)u(\bx)=0 \quad {\rm with} \quad \Omega = \frac{\omega l}{\hat{c}_{0}}
\label{eq:2dnormal}
\eeq
 and $\hat{c}_0=\sqrt{\hat{a}_0/\hat{\rho}_0}$. 
 
\indent We now introduce independent short- and long-scales $l$ and
$L$, and the ratio of these two scales $\epsilon=l/L \ll 1$
will provide the small (positive) parameter, $\epsilon$, to be used further on in
our homogenisation method. The two disparate length scales then
motivate new sets of dimensionless coordinates, namely ${\bf X}=\bx/L$ and $\bxi=\bx/l$. We augment these with a third 
set $\bzeta$, which are related to $\bxi$ via the
relations \eqref{eq:gen_coord}. It will be convenient to
use the orthogonal short-scale $\bxi$ for asymptotic expansions of (\ref{eq:gov_eqn}), and then revert to the lattice coordinates $\bzeta$ when we wish to impose the periodicity
conditions and later for performing integrals over the cell. As is conventional in multiple scale analysis we treat $\bxi, {\bf X}$ as being independent.

\indent We proceed by placing the disparate orthogonal coordinates into \eqref{eq:2dnormal}, noting that the periodicity of the functions $a$ and $\rho$ is specified on the short-scale only:
\ba
&\nabla_{\bxi}\cdot[a(\bxi)\nabla_{\bxi}u({\bf
      X},\bxi)]+\Omega^2\rho(\bxi) u ({\bf X},\bxi)
  \nonumber\\
  &\qquad +\epsilon[2a(\bxi)\nabla_\bxi
    +\nabla_{\bxi}a(\bxi)]\cdot\nabla_{{\bf X}}u({\bf X},\bxi)
  +\epsilon^2 a(\bxi)\nabla^2_{\bf X}u({\bf X},\bxi)=0.
\label{eq:two-scales2D}
\ea
Note that $u$ is now a function of independent separated-scale coordinates so $u=u ({\bf X},\bxi)$. Our focus will be on
analysing the motion of the system near specified eigenfrequencies. An
important point is that, dissimilar to orthogonal geometries, standing wave
eigenmodes exist for non-orthogonal geometries that do not necessarily satisfy in-phase or out-of-phase
boundary conditions at the edges of the cell. Notably for these
eigenmodes the phase-shift across the cell is complex and this in turn
gives rise to displacements having a non-zero imaginary component. As
a result our asymptotic method is no longer restricted to dealing with
real eigenmodes and we can analyse eigenfrequencies across the entirety of the Bloch spectrum and thereby perturb about any point in wavevector space. The Bloch conditions on the short-scale are applied using the general coordinates:
\beq
u|_{\zeta_{i}=1}=\exp\left(2i{\boldsymbol\kappa}\cdot{\bf e}_i \right) u|_{\zeta_{i}=-1} \quad {\rm and} \quad u_{,\zeta_i}|_{\zeta_{j}=1}=\exp\left(2i{\boldsymbol\kappa}\cdot{\bf e}_i \right) u_{,\zeta_i}|_{\zeta_{j}=-1}
\label{eq:periodicBC}
\eeq
where $u_{,\zeta_i}$ denotes partial differentiation with respect to $\zeta_i$. 
We now take the following ansatz
\beq
u({\bf X},\bxi)=u_0({\bf X},\bxi)+\epsilon u_1({\bf X},\bxi)+\epsilon^2 u_2({\bf X},\bxi)+\ldots, \quad \Omega^2=\Omega_0^2+\epsilon \Omega_1^2+\epsilon^2 \Omega_2^2+\ldots
\label{eq:expansion2D}
\eeq
which leads to a hierarchy of equations at different orders of $\epsilon$. The first three orders yield
\begin{subequations} 
\begin{equation}
(au_{0,\xi_i})_{,\xi_i} + \Omega_0^2 \rho u_{0}=0
\label{eq:leadingOrder}
\end{equation}
\vspace{-0.8cm}
\begin{equation}
(au_{1,\xi_i})_{,\xi_i} + \Omega_0^2\rho u_1=
-(2au_{0,\xi_i}+a_{,\xi_i}u_0)_{,X_i}
-\Omega_1^2 \rho u_0 
\label{eq:firstOrder}
\end{equation}
\vspace{-0.65cm}
\begin{equation}
\begin{split}
  (au_{2,\xi_i})_{,\xi_i} + \Omega_0^2\rho u_2=-au_{0,X_iX_i}
   -(2au_{1,\xi_i} + a_{,\xi_i}u_1)_{,X_i}\\
  -\Omega_1^2\rho u_1
   -\Omega_2^2\rho u_0.
\label{eq:secondOrder}
\end{split}
\end{equation} 
\end{subequations}
For the leading order equation there is a solution precisely at $\Omega_0$ and a corresponding eigenmode
$U_0(\bxi;\Omega_0)=U_0(\bzeta;\Omega_0)$,  with a fixed phase shift in $\bzeta$,
\beq
 u_0({\bf X},\bxi)=f_0({\bf X}) U_0(\bzeta;\Omega_0),
\label{eq:LeadingSolution}
\eeq where the the PDE governing $f_0({\bf X})$
 is to be determined. 

\indent We now use the perturbation method about the
frequency $\Omega_0$, and initially concentrate on the 
treatment of isolated eigenvalues. Repeated eigenvalues can and frequently do 
arise, and such cases require modifications. These cases are
often associated with Dirac-like cones and will be dealt with later. We now assume that there are $N$ inclusions per unit cell, hence the total surface $S=S_1-\left(S_2 \cup S_3\cup ...\cup S_{N+1} \right)$, where $S_1$ denotes the surface of the unit cell without holes and $S_j$ ($j=2\rightarrow N+1$) represent the surfaces of the holes. We impose Dirichlet conditions on the boundaries $\partial
S_j$ and deduce
\beq
u({\bf X},\bzeta)|_{\partial S_j}=0 \iff u_i({\bf X},\bzeta)|_{\partial S_j}=0, \quad i\in\ \mathbb{N}.
\label{eq:dirichletConditions}
\eeq
As these conditions are set in the short-scale $\bzeta$ we obtain
for $i=0$, $U_0(\bzeta;\Omega_0)|_{\partial S_j}=0$.

\indent To proceed we apply the Fredholm alternative. We first multiply equation \eqref{eq:firstOrder} by
$U_0^* $ and integrate over the cell's surface:
\ba
   & \iint_S \left( U_0^*(a u_{1,\xi_i})_{,\xi_i} +\Omega_0^2\rho
      U_0^* u_1\right) dS \nonumber\\
  &\qquad\qquad =-f_{0,X_i} \iint_S\left[(a
    |U_0|^2)_{,\xi_i}+\beta_i\right] dS -f_0\Omega_1^2 \iint_S\rho |U_0|^2 dS ,
  \label{eq:intermed1}
\ea 
\ba \text{where} \quad
\beta_i=a\left[U_0^* U_{0,\xi_i}-U_0 U_{0,\xi_i}^* \right],
\label{eq:beta_term}\ea and $|U_0|$ denotes the modulus of the complex displacement.

\indent The first term on the right hand side of (\ref{eq:intermed1})
vanishes when we use the planar divergence theorem along with the Bloch conditions \eqref{eq:periodicBC} directed along $\bzeta$. 
We continue by subtracting the cell integral of the conjugated Helmholtz equation $(aU^*_{0,\xi_i})_{,\xi_i} + \rho \Omega_0^2 U^*_{0}=0$  multiplied by $u_1$ to obtain:
\beq
  \iint \left[U_0^*(a u_{1,\xi_i})_{,\xi_i}
      - u_1(a U_{0,\xi_i}^*)_{,\xi_i} \right] dS
    = -f_{0,X_i} \iint \beta_i dS-f_0\Omega_1^2 \iint \rho|U_0|^2 dS.
\label{eq:intermed2}
\eeq 
Using Green's theorem, the left side of equation (\ref{eq:intermed2}) becomes
\beq
 \int a(\bzeta)\left(U_0^*\frac{\partial u_1}{\partial
     n}-u_1\frac{\partial U_0^*}{\partial n}\right) ds,
\label{eq:intermed3}
\eeq
where the line integral in the first line is over $\partial S=\partial S_1 \cup\partial S_2\cup\partial S_3...\cup\partial S_{N+1}$ and we have assumed that $\rho,a \in \mathbb{R}$. In equation \eqref{eq:intermed3} we find that by applying the Bloch conditions \eqref{eq:periodicBC} the terms on the opposing boundaries of the cell cancel each other out, and with the help of the Dirichlet boundary conditions on the inclusions, the terms above go to zero. Hence we are left only with the terms on the right hand side of equation \eqref{eq:intermed2} and if the surface integral of $\beta_i \neq 0$ then we are left with a first-order governing equation for $f_0({\bf{X}})$:
\beq
T_{i}^{(1)} f_{0,X_i}-\Omega_1^2 f_{0}=0, \quad T_{i}^{(1)}=\frac{-\iint \beta_i dS}{\iint \rho |U_0|^2 dS},
\label{eq:T1D} \eeq for $i=1,2$. An additional point to note is that
if the integral of $\beta_i$ over the unit cell is non-zero then this
implies that the local variation along a certain path in
frequency-wavevector space is linear and as such the envelope
modulation, along this specified path, is governed by the above
equation. Another nuance is
that for certain geometries we may obtain points that are linear along
one approach in frequency-wavevector space  and quadratic (or higher
order) along another and in this case a $\beta_i$ term will propagate
to higher order. However along the path which has locally nonlinear
curvature the term involving $\beta_i$ vanishes, hence the $\beta_i$ term is neglected as we proceed to quadratic order.

\indent Assuming that the integral involving $\beta_i$ is zero, the simplified equation for $u_1$ then has an explicit solution
\beq
u_1({\bf X},\bxi)=f_1({\bf X}) U_0(\bxi;\Omega_0)
    +\nabla_{\bf X} f_0({\bf X})\cdot {\bf U_1}(\bxi).
\label{eq:u1Solution1}
\eeq Substituting into equation \eqref{eq:firstOrder}, gives a set of coupled equations to be solved for ${\bf U}_1$
\beq
(aU_{1j,\xi_i})_{,\xi_i} + \Omega_0^2\rho U_{1j}=
-2aU_{0,\xi_j}-a_{,\xi_j}U_0\quad\text{for}\quad j=1,2,
\label{eq:U1eqs}
\eeq  where the auxiliary function $U_{1j}$ also satisfies the Bloch conditions. 

\indent We now proceed to second order and find the effective equation governing the envelope
modulation by using similar solvability conditions to
those employed at the previous order. We multiply equation
\eqref{eq:secondOrder} by $U_0^*$,
subtract the product of the complex conjugate of equation \eqref{eq:leadingOrder} with $u_2/f_0^*$ and 
integrate over the unit cell, thereby
eventually giving us a partial differential equation purely on the long-scale
\beq
 T_{ij}^{(2)} f_{0,X_iX_j}+\Omega_2^2 f_0=0, \quad \text{with}\quad
  T_{ij}^{(2)}=\frac{t_{ij}}{\iint \rho |U_0|^2 dS} \quad \text{for} \quad i,j=1,2.
\label{eq:TijEquation}
\eeq
The coefficients $T_{ij}^{(2)}$ encode the short-scale
behaviour of our effective medium within the purely long-scale
governing equation and the $t_{ij}$'s are given as
\beq
t_{ii}=\iint a|U_0|^2dS+2\iint aU_{1_i,\xi_i}U_0^*dS+\iint a_{,\xi_i}U_{1_i}U_0^*dS,
\label{eq:t11}
\eeq
\vspace{-0.5cm}
\beq
t_{ij}=2\iint aU_{1_j,\xi_i}U_0^*dS+\iint a_{,\xi_i}U_{1_j}U_0^*dS \quad {\rm for} \quad i\neq j.
\label{eq:tij}
\eeq  As we desired, we are finally left with an effective homogenised
equation (\ref{eq:TijEquation}) to be solved for $f_0({\bf X})$.

\subsection{Kirchhoff-Love equation}
\indent Herein we shall consider the simplified framework of the
Kirchhoff-Love plate theory that allows for bending moments and
transverse shear forces. The resulting PDE is fourth-order in space
and second-order in time (although we shall consider the time-harmonic
problem). This simplified model for flexural waves is due to a
relationship between the stiffness/thickness of the plate and, as in (\ref{eq:kirchoff}), is given explicitly as
\beq
\nabla_{\bf x}^2\left[\beta\left(\bx \right)\nabla_{\bf x}^2 u\left(\bx \right) \right]-\mu\left(\bx \right){\hat \Omega}^2 u\left(\bx \right) = 0.
\label{eq:2DPrimeK}
\eeq 
Notably both material parameters $\beta$ and $\mu$,
similar to the material parameters $\hat\rho, \hat a$ in the previous section, have periodic boundary conditions on opposite sides of the unit cell. Henceforth we operate in the non-dimensional setting and drop the hat decoration. The short and long-scale coordinates are identical to those used in the prior section, where $\epsilon$ is once again defined by the ratio of the length scales.

\indent We reiterate that we can perturb about any point in wavevector space hence the Bloch conditions on the short-scale, applied using the general coordinates, are given explicitly as 
\[u|_{\zeta_{m}=1}=\exp\left(2i{\boldsymbol\kappa}\cdot{\bf e}_i \right) u|_{\zeta_{m}=-1}, \quad  u_{,\zeta_i}|_{\zeta_{m}=1}=\exp\left(2i{\boldsymbol\kappa}\cdot{\bf e}_i \right) u_{,\zeta_i}|_{\zeta_{m}=-1}\]
\vspace{-0.6cm}
\beq
u_{,\zeta_i \zeta_j}|_{\zeta_{m}=1}=\exp\left(2i{\boldsymbol\kappa}\cdot{\bf e}_i \right) u_{,\zeta_i \zeta_j}|_{\zeta_{m}=-1}, \quad u_{,\zeta_i \zeta_j \zeta_k}|_{\zeta_{m}=1}=\exp\left(2i{\boldsymbol\kappa}\cdot{\bf e}_i \right) u_{,\zeta_i \zeta_j \zeta_k}|_{\zeta_{m}=-1}
\label{eq:periodicBCK}
\eeq
where $u_{,\zeta_i}$ denotes partial differentiation with respect to $\zeta_i$; similarly $u_{,\zeta_i \zeta_j}$ and $u_{,\zeta_i \zeta_j \zeta_k}$ denote the second and third order partial differentiations with respect to $\zeta_i, \zeta_j$ and $\zeta_i, \zeta_j,\zeta_k$, respectively. Recall that repeated indices denote summation. Similar to the previous section, the array of holes on the plate have homogeneous Dirichlet conditions imposed on their boundary, in addition to Neumann conditions, $u=\partial u/\partial r=0$.

\indent We separate ${\bf x}$ into the two disparate length scales and expand out the displacement and frequency terms accordingly to eventually obtain
\beq
(\beta u_{0,\xi_i\xi_i})_{,\xi_j\xi_j} - \mu\Omega_0^2 u_0= 0,
\label{eq:leadingOrder2DK}
\eeq
\vspace{-0.6cm}
\beq
(\beta u_{1,\xi_i\xi_i})_{,\xi_j\xi_j}+2(\beta u_{0,\xi_i\xi_i})_{,X_j \xi_j}+ 2(\beta u_{0,X_i\xi_i})_{,\xi_j\xi_j} - \mu \Omega_0^2 u_1 - \mu \Omega_1^2 u_0 = 0,
\label{eq:FirstOrder2DK}
\eeq
\vspace{-0.6cm}
\begin{align}
&(\beta u_{2,\xi_i\xi_i})_{,\xi_j\xi_j}+2(\beta u_{1,\xi_i\xi_i})_{,X_j \xi_j} + 2(\beta u_{1,X_i \xi_i})_{,\xi_j\xi_j} + (\beta u_{0,\xi_i\xi_i})_{,X_jX_j}+
\nonumber\\
&\qquad4(\beta u_{0,X_i \xi_i})_{,X_j \xi_j} +(\beta u_{0,X_iX_i})_{,\xi_j\xi_j} - \mu \Omega_0^2 u_2 -\mu \Omega_1^2 u_1 -\mu \Omega_2^2 u_0 = 0.
\label{eq:SecondOrder2DK}
\end{align} Note that despite the theory being generalised for non-orthogonal geometries we shall once again opt to leave our equations in the orthogonal system, for succinctness.

\indent The leading order problem is independent of the long-scale, hence the associated displacement can be written as $u_0=f_0({\bf X})U_0({\boldsymbol \xi};\Omega_0)=f_0({\bf X})U_0({\boldsymbol \zeta};\Omega_0)$, where $U_0 \in \mathbb{C}$ and it satisfies the following equation
\beq
(\beta U_{0,\xi_i\xi_i})_{,\xi_j\xi_j} - \mu\Omega_0^2 U_0= 0.
\label{eq:NewleadingOrder2DK}
\eeq We now integrate over the cell the difference between the product of equation (\ref{eq:FirstOrder2DK}) and $U_0^*$ and the product of the complex conjugate of equation (\ref{eq:FirstOrder2DK}) and $u_1/f_0$ to obtain the following,
\begin{align}
&\iint ((\beta u_{1,\xi_i\xi_i})_{,\xi_j\xi_j}U_0^*-(\beta U_{0,\xi_i\xi_i})_{,\xi_j\xi_j}u_1)dS + 
\nonumber\\
&\quad2\iint ((\beta u_{0,\xi_i\xi_i})_{,\xi_jX_j}U_0^*+(\beta u_{0,\xi_iX_i})_{,\xi_j\xi_j}U_0^*)dS-\iint \mu \Omega_1^2 u_0 U_0^* dS=0.
\label{eq:interm1rst2DK}
\end{align}
Recall that the transformation from the orthogonal coordinates $\bxi$ to $\bzeta$ is linear, therefore $u_{\xi_i}, u_{\xi_i,\xi_j}$ equate to a linear combination of $u_{\zeta_i}, u_{\zeta_i, \zeta_j}$ terms, respectively. 	Using integration by parts and the Bloch conditions stated in $\bzeta$ space \eqref{eq:periodicBC}, the first integral term in equation \eqref{eq:interm1rst2DK} vanishes. 

\indent After successive integration by parts the second integral term of equation (\ref{eq:interm1rst2DK}) eventually cancels down to the following 
\beq
\iint \eta_j dS, \hspace{0.4cm} \eta_j= 2\beta\left(U_{0,\xi_j}U_{0,\xi_i\xi_i}^*-U_{0,\xi_j}^*U_{0,\xi_i\xi_i} \right).
\label{eq:beta_termK}\eeq
The above term \eqref{eq:beta_termK}, is analogous to the $\beta_i$
term found in equation \eqref{eq:beta_term}, therefore for locally
non-linear curvature the above term integrates to zero which in turn
implies that $\Omega_1=0$ and we proceed to quadratic order. 

If, however, the above integral is non-zero we obtain the following first-order effective equation
\beq
 T_{i}^{(1)} f_{0,X_i}-\Omega_1^2 f_0=0, \quad T_{i}^{(1)}=\frac{\iint \eta_i dS}{\iint \mu |U_0|^2 dS} ,
\label{eq:T1DK} \eeq
 where $\eta_i$ is defined in \eqref{eq:beta_termK}. 

Proceeding with the assumption that $\Omega_1=0$, inserting this
result into equation (\ref{eq:FirstOrder2DK}) and solving for
$u_1(X,\bxi)$ gives:
\beq
u_1({\bf X},{\boldsymbol \xi})=f_1({\bf X}) U_0(\bxi;\Omega_0)+\nabla_{\bf X} f_0({\bf X})\cdot {\bf U_1}(\bxi).
\label{eq:u1SolutionK}
\eeq The homogeneous component of the above solution $f_1({\bf X})U_0(\bxi;\Omega_0)$ is absorbed by the leading order solution, whilst the equation to solve for the inhomogeneous component is,
\beq
(\beta U_{1k,\xi_i\xi_i})_{,\xi_j\xi_j} - \mu \Omega_0^2 U_{1k} = -(2(\beta U_{0,\xi_i\xi_i})_{,\xi_k}+ 2(\beta U_{0,\xi_k})_{,\xi_i\xi_i}).
\label{eq:u1EquationK}
\eeq Note that $U_{1k}$ must respect the boundary conditions, stated previously in equations (\ref{eq:periodicBC}), and the homogeneous conditions on the inclusions.

\indent We now turn to the second order equation (\ref{eq:SecondOrder2DK}). We multiply equation (\ref{eq:SecondOrder2DK}) by $U_0^*$ and subtract the product of the complex conjugate of equation (\ref{eq:leadingOrder2DK}) by $u_2/f_0$ and then integrate over the elementary cell,
\begin{align}
&\iint \left(U_0^*(\beta u_{2,\xi_i\xi_i})_{,\xi_j\xi_j}-u_2(\beta U_{0,\xi_i\xi_i})_{,\xi_j\xi_j}\right)dS +
\nonumber\\
&\quad
\iint 2U_0^*\left((\beta u_{1,\xi_i\xi_i})_{,\xi_jX_j}+(\beta u_{1,X_i\xi_i})_{,\xi_j\xi_j}\right)dS +
\nonumber\\
&\qquad
\iint U_0^*\left((\beta u_{0,\xi_i\xi_i})_{,X_jX_j}+4(\beta u_{0,\xi_iX_i})_{,\xi_jX_j}+(\beta u_{0,X_iX_i})_{,\xi_j\xi_j}\right)dS+
\nonumber\\
&\quad\qquad
-\iint \mu \Omega_2^2u_0U_0 ^*dS = 0.
\label{eq:u2IntermediateK}
\end{align}
The first integral is nigh on identical to the integral found in equation \eqref{eq:interm1rst2DK}  and equates to zero in a similar manner. The second integral term is separated into two parts by splitting $u_1$ into its homogeneous and inhomogeneous components. The homogeneous term accompanying $f_1$ integrates to zero when $\eta_i=0$ \eqref{eq:beta_termK}; after some algebra we are left with an equation of the form
\beq
T_{ij}f_{0,X_iX_j}-\Omega_2^2 f_0=0,\quad
  T_{ij}^{(2)}=\frac{t_{ij}}{\iint \mu|U_0|^2 dS},
\label{eq:TijEquationK}
\eeq where the $T_{ij}^{(2)}$'s are given explicitly as
\beq
t_{ij}=\hat{t}_{ij} \quad {\rm for} \quad i\neq j, \quad \text{\rm where} 
\nonumber\eeq
\vspace{-0.5cm}
\beq
\hat{t}_{ij}=\iint U_0^* \left(2\left[(\beta U_{1i,\xi_k\xi_k})_{,\xi_j}+(\beta U_{1i,\xi_j})_{,\xi_k\xi_k} \right] + 4(\beta U_{0,\xi_i})_{,\xi_j} \right) dS,
\label{eq:tijK}
\eeq
\beq
{\rm and} \quad t_{ii}=\hat{t}_{ii}+ \iint U_0^* \left(	\beta U_{0,\xi_i\xi_i}+(\beta U_0)_{,\xi_i\xi_i}\right)dS.
\label{eq:tiiK}
\eeq

\section{Exact solutions for constrained points}
\label{sec:exact}
In the following section we specify to a doubly-periodic array of constrained points. This configuration is useful as for both systems we are then able to obtain precise analytical solutions using Fourier series. This property, in conjunction with the prior asymptotics, is used to obtain explicit representations of the coefficients, $T^{(1)}_i,T^{(2)}_{ij}$ in equations \eqref{eq:T1D}, \eqref{eq:TijEquation}, \eqref{eq:T1DK}, \eqref{eq:TijEquationK}. 

\subsection{The Helmholtz equation} 
In the case of constrained points, we augment the Helmholtz equation with a term representing the reaction forces induced by each one:
\beq
\left(\nabla_{\bx}^2+\Omega^2 \right)u({\bx})=\sum\limits_{n,m} \sum\limits_{k=1}^N F_{n,m}^k \delta\left({\bx}-{\bf I}_{n,m}^k \right),
\label{eq:simple_supports_eqn}\eeq where we have assumed for simplicity that $a=\rho=1$. Here $n,m \in \mathbb{Z}$, $F_{n,m}^k, {\bf I}_{n,m}^k$ denote the force and position associated with the $k$'th inclusion in the $(n,m)$'th cell, and
\beq
\delta\left({\bx}-{\mbox{\boldmath$\alpha$}}\right)=\delta\left(x_1-\mbox{$\alpha$}_1 \right)\delta\left(x_2- \mbox{$\alpha$}_2 \right).
\label{eq:delta_prop1} \eeq In general there are $N$ inclusions per unit cell, located at ${\bf I}_{n,m}^k={\bf I}_{n,m}+{\bf I}^k$, where ${\bf I}_{n,m}=2\left(n{\bf e}_1 + m{\bf e}_2 \right)$ specifies the cell and ${\bf I}^k$ identifies the location of the $k$'th inclusion within the cell.

Using the prior asymptotics to inform this section, we deduce, after some algebra, the following leading order equation:
\beq
\left(\nabla_{\bxi}^2+\Omega_0^2 \right)U_0({\bxi})= \sum\limits_{k=1}^N F_{0}^k \delta\left({\bxi}-{\bf I}^k \right).
\label{eq:leadingorder_FT} \eeq The leading order displacement is given by $u_0\left(\bxi,\bX \right)=U_0(\bxi)f_0(\bX)$, and the forcing term $F_0^k$ is related to reaction forces $F_{n,m}^k$ via the following periodicity condition and expansion:
\beq
F_{n,m}^k= \exp \left[i\left(\boldsymbol{\kappa}\cdot{\bf I}_{n-\hat{n},m-\hat{m}} \right) \right] F_{\hat{n},\hat{m}}^k, \hspace{0.4cm} F_{\hat{n},\hat{m}}^k=f_0(\bX)F_0^k + \epsilon\hat{F}_1^k(\bX)+\epsilon^2\hat{F}_2^k(\bX)+... \hspace{0.1cm}.
\label{eq:forcing_relations} \eeq Here we denote $\mbox{\boldmath$\kappa$}=(\kappa_1,\kappa_2)$, and the $(\hat{n},\hat{m})$'th cell is taken as an arbitrary reference cell in which equation \eqref{eq:leadingorder_FT} is valid. Due to the periodic arrangement of the inclusions, the displacement response can be written as
\beq
U_0(\bxi)=\sum\limits_{{\bf G}} \hat{U}_0({\bf G}) \exp\left[i \left({\bf G}-{\bkappa}\right)\cdot{\bxi} \right],
\label{eq:U0_Bloch_Theorem} \eeq where ${\bf G}$ is the reciprocal lattice vector defined via
\beq
{\bf G}=n\hat{\bf e}_1+m\hat{\bf e}_2, \hspace{0.4cm} {\bf e}_i\cdot\hat{\bf e}_j=\pi \delta_{ij}, \hspace{0.4cm} n, m \in \mathbb{Z},
\label{eq:recip_lattice_defn} \eeq so from  \eqref{eq:gen_recip_vec} we have $\hat{\bf e}_j={\bf e}_j^*/2$. We substitute \eqref{eq:U0_Bloch_Theorem}  into \eqref{eq:leadingorder_FT} and multiply through by $\exp\left[-i\left({\bf G'}-{\bkappa} \right) \right]$, where ${\bf G'}$ is a fixed reciprocal lattice vector, to give us
\beq
\begin{split}
\sum\limits_{{\bf G}} \left(-|{\bf G}-{\bkappa}|^2+\Omega_0^2\right)\hat{U}_0({\bf G}) \exp\left[i\left({\bf G}-{\bf G'}\right)\cdot{\bxi}\right]\\
=\sum\limits_{k=1}^N F_{0}^k \delta\left({\bxi}-{\bf I}^k \right) \exp\left[-i\left({\bf G'}-{\bkappa}\right)\cdot{\bxi}\right].
\label{eq:solving_leading_orderFT2}
\end{split}
\eeq Subsequently we integrate over the elementary cell to obtain the following expression for the short-scale displacement component:   
\beq
U_0({\bxi})=-\frac{1}{A}\sum\limits_{{\bf G}} \sum\limits_{k=1}^N \frac{F_{0}^k \exp\left[i \left({\bf I}^k-{\bxi} \right)\cdot\left({\bf G}-\mbox{\boldmath$\kappa$}\right) \right]}{|{\bf G}-\mbox{\boldmath$\kappa$}|^2-\Omega_0^2},
\label{eq:exact_displ} \eeq where $A=4|{\bf e}_1 \times {\bf e}_2|$ is the area of the unit cell. Enforcing the Dirichlet condition $U_0(I^k)=0$ at the simple supports gives us the following $N$ equations:
\beq
U_0\left({\bf I}^b \right)=\sum\limits_{k=1}^N \varrho_{b,k} F_{0}^k=0, \hspace{0.3cm}  \varrho_{\alpha,\beta}=\sum\limits_{n,m}\frac{\exp\left[i \left({\bf I}^\alpha-{\bf I}^\beta \right)\cdot\left({\bf G}- \mbox{\boldmath$\kappa$}\right) \right]}{|{\bf G}-\mbox{\boldmath$\kappa$}|^2-\Omega_0^2},
\label{eq:constrained_pt_disp_eqn}\eeq where $b=1 \rightarrow N$. The dispersion relation easily follows
as det$(\boldsymbol{\varrho})=0$, where $\boldsymbol{\varrho}$ is the matrix with elements equal to $\varrho_{i,j}$. 

\indent The above dispersion relation clearly contains singularities, and these correspond to solutions for  waves propagating in a homogeneous medium without
constraints. It was previously shown \cite{antonakakis12a} for square lattices that on occasion the corresponding solutions also satisfy the Dirichlet condition at the centre of the square cell, and hence lie on the dispersion curves for the pinned structure. This observation extends to all Bravais lattices, and when this perfect solution coincides with a dispersion curve we can deduce the leading order displacement with ease. We find that this occurs regularly at multiple crossing points for various Bravais lattices (Fig. \ref{fig:tri_dispersion_total2}, Fig. \ref{fig:honey_dispersion_total2}), resulting in so-called generalised Dirac points. Using this property, along with the leading order displacement \eqref{eq:exact_displ}, we find for $N=1$ that the solution at these generalised Dirac points is formed from a linear combination of 
\beq
\sin\left[\left({\bf G}- \mbox{\boldmath$\kappa$}\right)\cdot{\bxi}\right], \hspace{0.2cm}\sin\left[\left({\bf G}- \mbox{\boldmath$\kappa$}\right)\cdot\hat{\bxi}\right],\hspace{0.2cm} \cos\left[\left({\bf G}- \mbox{\boldmath$\kappa$}\right)\cdot{\bxi}\right] -\cos\left[\left({\bf G}- \mbox{\boldmath$\kappa$}\right)\cdot\hat{\bxi}\right],
\label{eq:constr_pt_brav_perfect}\eeq where $\hat{\bxi}=\left(\xi_1,-\xi_2 \right)$. These solutions satisfy the Helmholtz equation, $\left(\nabla_{\bxi}^2+\Omega^2 \right)u({\bxi})=0$ and the constraint at the origin of the cell. The multiplicity, for fixed $\Omega$, is dependent on the number of linearly independent solutions formed from the functions \eqref{eq:constr_pt_brav_perfect}. For example, the hexagonal lattice has a Dirac point at $\Omega^2\left(\Gamma \right)=|{\bf G}|^2=(4/3)\pi^2$ where $(n,m)=(1,0),(0,1)$ in equation \eqref{eq:recip_lattice_defn}, so the leading order solution is found to be
\[ u_0=f^{(1)}\left({\bf X}\right)\sin\left(\nu_{-}\right)+f^{(2)}\left({\bf X}\right)\sin\left(\nu_{+} \right)+f^{(3)}\left({\bf X}\right)\sin\left(\tau \right)+f^{(4)}\left({\bf X}\right)\left[\cos\left(\nu_{+} \right)-\cos\left(\tau\right) \right]\]
\vspace{-0.5cm}
\beq
+f^{(5)}\left({\bf X}\right)\left[\cos\left(\nu_{-}\right)-\cos\left(\tau\right)\right], \hspace{0.2cm} \text{where} \hspace{0.1cm} \nu_{\pm}=\pi\left(\xi_1 \pm \xi_2/\sqrt{3}\right), \hspace{0.1cm} \tau=\left(2\pi/\sqrt{3}\right)\xi_2.
\label{eq:dirac_hex_gamma}\eeq A similar method can be used at other points in the Bloch diagram and for different periodic structures.

\indent For isolated eigenvalues that give non-singular solutions to the dispersion relation \eqref{eq:constrained_pt_disp_eqn}, we use the Fourier series \eqref{eq:exact_displ} to obtain the precise asymptotics. If the local curvature at a fixed point in wavevector space is linear, we find that the non-zero first-order coefficient found in equation  \eqref{eq:T1D} is given explicitly by
\[T_{i}^{(1)}=2i\frac{\Lambda_{i}^{(1)}}{\Upsilon},\hspace{0.3cm} \text{where} \hspace{0.3cm} \Upsilon=\frac{1}{A^2}\sum\limits_{{\bf G}} \sum_{\substack{k,l=1 \\ k \neq l}}^N
 \frac{F_{0}^k \gamma_{k,l}}{\left(|{\bf G}-\mbox{\boldmath$\kappa$}|^2-\Omega_0^2 \right)^2},\]
\begin{align}
\Lambda_{i}^{(1)}=\frac{1}{A^2}\sum\limits_{{\bf G}} \sum_{\substack{k,l=1 \\ k \neq l}}^N \frac{\left[({\bf G})_i-\kappa_i \right]\left(F_{0}^k \gamma_{k,l} \right)}{\left(|{\bf G}-\mbox{\boldmath$\kappa$}|^2-\Omega_0^2 \right)^2}, 
\\ 
\gamma_{l,k}=F_{0}^k+ F_{0}^l \exp\left[i\left({\bf I}^k-{\bf I}^l\right)\cdot\left({\bf G}-\mbox{\boldmath$\kappa$}\right) \right],
\label{eq:gen_Ti1}
\end{align} and $({\bf G})_i $ is the $i$'th component of the reciprocal lattice vector given by \eqref{eq:recip_lattice_defn}. If $T_{i}^{(1)}=0$ we deduce that $\Omega_1=0$ and so seek a solution corresponding to \eqref{eq:u1Solution1} for the first order displacement. With in mind we assume that the first-order reaction force, \eqref{eq:forcing_relations}, takes the form 
\beq
\hat{F}_1^k({\bf X})=\nabla_{\bX}f_0({\bX})\cdot{\bf F}_1^k+f_1({\bX})F_0^k,
\label{eq:first_order_reaction_force} \eeq where ${\bf F}_1^k=\left(F_{11}^k, F_{12}^k \right)$ is to be found. The governing equation for $U_{1j}$, \eqref{eq:U1eqs} , is augmented with the forcing term $\sum\limits_{k}F_{1j}^k \delta \left({\bxi}-{\bf I}^k \right)$ and solved accordingly:
\beq
U_{1j}({\bxi})=\frac{1}{A}\sum\limits_{{\bf G}} \sum\limits_{k=1}^N\frac{\exp\left[i \left({\bf I}^k-{\bxi} \right)\cdot\left({\bf G}- \mbox{\boldmath$\kappa$}\right)\right]}{|{\bf G}-\mbox{\boldmath$\kappa$}|^2-\Omega_0^2 } \left(2i\frac{F_0^k \left[({\bf G})_j-\kappa_j\right]}{|{\bf G}-\mbox{\boldmath$\kappa$}|^2-\Omega_0^2 }-F_{1j}^k \right).
\label{eq:exact_displ_firstorder} \eeq ${\bf F}_1^k$ is chosen to ensure that the Dirichlet condition is satisfied by ${\bf U}_1\left({\bxi}\right)$; in the case that $N=1$ this implies ${\bf F}_1^k={\bf 0}$.

\indent We can now extract the $T_{ij}^{(2)}$ values in equation \eqref{eq:TijEquation}, integrating the necessary terms by hand. Substituting the leading and first order displacements \eqref{eq:exact_displ} and \eqref{eq:exact_displ_firstorder} into equations \eqref{eq:t11} and \eqref{eq:tij} we get
\vspace{-0.5cm}
\[T_{ii}^{(2)}=1+2\frac{\Lambda_{ii}^{(2)}}{\Upsilon}, \hspace{0.3cm} T_{ij}^{(2)}=2\frac{\Lambda_{ij}^{(2)}}{\Upsilon}, \hspace{0.3cm} \text{where} \]
\vspace{-0.5cm}
\beq
\begin{split}
\Lambda_{ij}^{(2)}=-\frac{1}{A^2}\sum\limits_{{\bf G}} \sum_{\substack{k,l=1 \\ k \neq l}}^N \frac{\left[({\bf G})_i-\kappa_i \right]\gamma_{k,l}}{\left(|{\bf G}-\mbox{\boldmath$\kappa$}|^2-\Omega_0^2\right)^2} \left(iF_{1j}^k+\frac{2F_0^k\left[({\bf G})_j-\kappa_j \right]}{|{\bf G}-\mbox{\boldmath$\kappa$}|^2-\Omega_0^2}\right).
\label{eq:gen_Tij}
\end{split}
\eeq Note that for orthogonal geometries, such as the square and rectangular Bravais lattices, the cross-derivative coefficients, $T_{ij}^{(2)}$ for $i \neq j$, are equal to zero.


\subsection{Kirchhoff-Love equation}
We now consider the case of pinned platonic crystals (PPCs), and once again deduce an exact dispersion relation using Fourier series. The material parameters $\beta$, $\mu$ are set to unity and the support boundary conditions are subsumed into the Kirchhoff-Love plate equation:
\beq
\left(\nabla_{\bx}^4-\Omega^2 \right)u({\bx})=\sum\limits_{n,m} \sum\limits_{k=1}^N F_{n,m}^k \delta\left({\bx}-{\bf I}_{n,m}^k \right).
\label{eq:simple_supports_eqnK}\eeq  Analogously to the previous section, we deduce the leading order problem
\beq
\left(\nabla_{\bxi}^4-\Omega_0^2 \right)U_0({\bxi})= \sum\limits_{k=1}^N F_{0}^k \delta\left({\bxi}-{\bf I}^k \right),
\label{eq:leadingorder_FTK} \eeq and obtain the following leading order solution:
\beq
U_0({\bxi})=\frac{1}{A}\sum\limits_{{\bf G}} \sum\limits_{k=1}^N \frac{F_{0}^k \exp\left[i \left({\bf I}^k-{\bxi} \right)\cdot\left({\bf G}-\mbox{\boldmath$\kappa$}\right) \right]}{|{\bf G}-\mbox{\boldmath$\kappa$}|^4-\Omega_0^2},
\label{eq:exact_displK} \eeq where once again $u_0\left(\bxi,\bX \right)=U_0(\bxi)f_0(\bX)$. Enforcing the Dirichlet condition $U_0(I^k)=0$ at the point supports gives us the following matrix equation:
\beq
U_0\left({\bf I}^b \right)=\sum\limits_{k=1}^N \varrho_{b,k} F_{0}^k=0, \hspace{0.3cm}  \varrho_{\alpha,\beta}=\sum\limits_{n,m}\frac{\exp\left[i \left({\bf I}^\alpha-{\bf I}^\beta \right)\cdot\left({\bf G}- \mbox{\boldmath$\kappa$}\right) \right]}{|{\bf G}-\mbox{\boldmath$\kappa$}|^4-\Omega_0^2},
\label{eq:constrained_pt_disp_eqnK}\eeq and once again the
dispersion relation is  given by det$(\boldsymbol{\varrho})=0$. It was shown by \cite{evans07a} that the fundamental Green's function of the thin-plate equation has a vanishing first derivative as one approaches the location of the source. Therefore it follows for our system that only the Dirichlet condition needs to be imposed at the location of the inclusions, as the Neumann condition is automatically satisfied in the case of zero-radius holes.

\indent An analogous formula to \eqref{eq:gen_Ti1} for the $T_{i}^{(1)}$ coefficients for PPCs is given by
\[T_{i}^{(1)}=4i\frac{\Lambda_{i}^{(1)}}{\Upsilon},\hspace{0.3cm} \text{where} \hspace{0.3cm} \Upsilon=\frac{1}{A^2}\sum\limits_{{\bf G}} \sum_{\substack{k,l=1 \\ k \neq l}}^N
 \frac{F_{0}^k \gamma_{k,l}}{\left(|{\bf G}-\mbox{\boldmath$\kappa$}|^4-\Omega_0^2 \right)^2},\]
\begin{align}
\Lambda_{i}^{(1)}=\frac{1}{A^2}\sum\limits_{{\bf G}} \sum_{\substack{k,l=1 \\ k \neq l}}^N \frac{\left[({\bf G})_i-\kappa_i \right]|{\bf G}-\mbox{\boldmath$\kappa$}|^2\left(F_{0}^k \gamma_{l,k} \right)}{\left(|{\bf G}-\mbox{\boldmath$\kappa$}|^4-\Omega_0^2 \right)^2}, 
\label{eq:gen_Ti1K}
\end{align} where the notation of \eqref{eq:gen_Ti1} is used again. In a similar manner to equation \eqref{eq:exact_displ_firstorder} there is an additional forcing term ${\bf F}_1^k=\left(F_{11}^k, F_{12}^k \right)$ to be found. The inhomogeneous component of equation \eqref{eq:u1SolutionK} is given by
\beq
U_{1j}({\bxi})=-\frac{1}{A}\sum\limits_{{\bf G}} \sum\limits_{k=1}^N\frac{\exp\left[i \left({\bf I}^k-{\bxi} \right)\cdot\left({\bf G}- \mbox{\boldmath$\kappa$}\right)\right]}{|{\bf G}-\mbox{\boldmath$\kappa$}|^4-\Omega_0^2 } \left(4i\frac{F_0^k \left[({\bf G})_j-\kappa_j\right]|{\bf G}-\mbox{\boldmath$\kappa$}|^2}{|{\bf G}-\mbox{\boldmath$\kappa$}|^4-\Omega_0^2 }-F_{1j}^k \right).
\label{eq:exact_displ_firstorderK} \eeq Again, ${\bf F}_1^k$ is chosen to ensure that the Dirichlet condition is satisfied. The $T_{ij}^{(2)}$ values are then given by 
\[T_{ij}^{(2)}=\frac{\Lambda_{ij}^{(2)}}{\Upsilon}, \hspace{0.3cm} T_{ii}^{(2)}=\frac{1}{\Upsilon}\left(\Lambda_{ii}^{(2)} -2\frac{F_{0}^k \gamma_{k,l}|{\bf G}-\mbox{\boldmath$\kappa$}|^2}{\left(|{\bf G}-\mbox{\boldmath$\kappa$}|^4-\Omega_0^2 \right)^2} \right), \hspace{0.3cm} \text{where} \]
\begin{align}
\Lambda_{ij}^{(2)}=\frac{1}{A}\sum\limits_{{\bf G}} \sum_{\substack{k,l=1 \\ k \neq l}}^N \frac{4\left[({\bf G})_j-\kappa_j \right] |{\bf G}-\mbox{\boldmath$\kappa$}|^2 \gamma_{k,l}}{\left(|{\bf G}-\mbox{\boldmath$\kappa$}|^4-\Omega_0^2\right)^2}\left(iF_{1j}^k+\frac{4F_0^k\left[({\bf G})_i-\kappa_i \right]|{\bf G}-\mbox{\boldmath$\kappa$}|^2}{|{\bf G}-\mbox{\boldmath$\kappa$}|^4-\Omega_0^2}\right) \nonumber \\
-\frac{4\left[({\bf G})_i-\kappa_i \right]\left[({\bf G})_j-\kappa_j \right]F_{0}^k \gamma_{k,l}}{\left(|{\bf G}-\mbox{\boldmath$\kappa$}|^4-\Omega_0^2\right)^2}.
\label{eq:gen_TijK}
\end{align} As was the case for the Helmholtz equation, for orthogonal geometries $T_{ij}^{(2)}=0$ for $i \neq j$ at standing wave frequencies.  

\section{Results}
\label{sec:results}
We are now in a position to compare the results of the asymptotic theory with the exact solutions, and also to investigate dynamic
anisotropic effects. We choose to focus on selected lattice geometries that offer novel
features, distinct from those of the previously analysed square geometry. For the Helmholtz equation, we consider rhombic and hexagonal lattices, and additionally the honeycomb lattice  due to its underlying relation to graphene. Note that the underlying periodicity of this structure is identical to that of the hexagonal lattice, and the two share an identical irreducible Brillouin zone. The edges of the
Brillouin zone for the hexagonal and
rhombic lattices can be found by halving the values in
\eqref{eq:gen_bdry_pts}; the parameter values are fixed as
$(\alpha,\theta)=\left(1,\pi/3 \right)$ and $(0.7,\arccos(0.35))$,
respectively. For the Kirchhoff-Love equation, for ease of computation, we solely examine the hexagonal lattice.

\subsection{The Helmholtz equation}
\subsubsection{Hexagonal lattice}
\begin{figure}[htb!]
\begin{center}
    \includegraphics[scale=0.7]{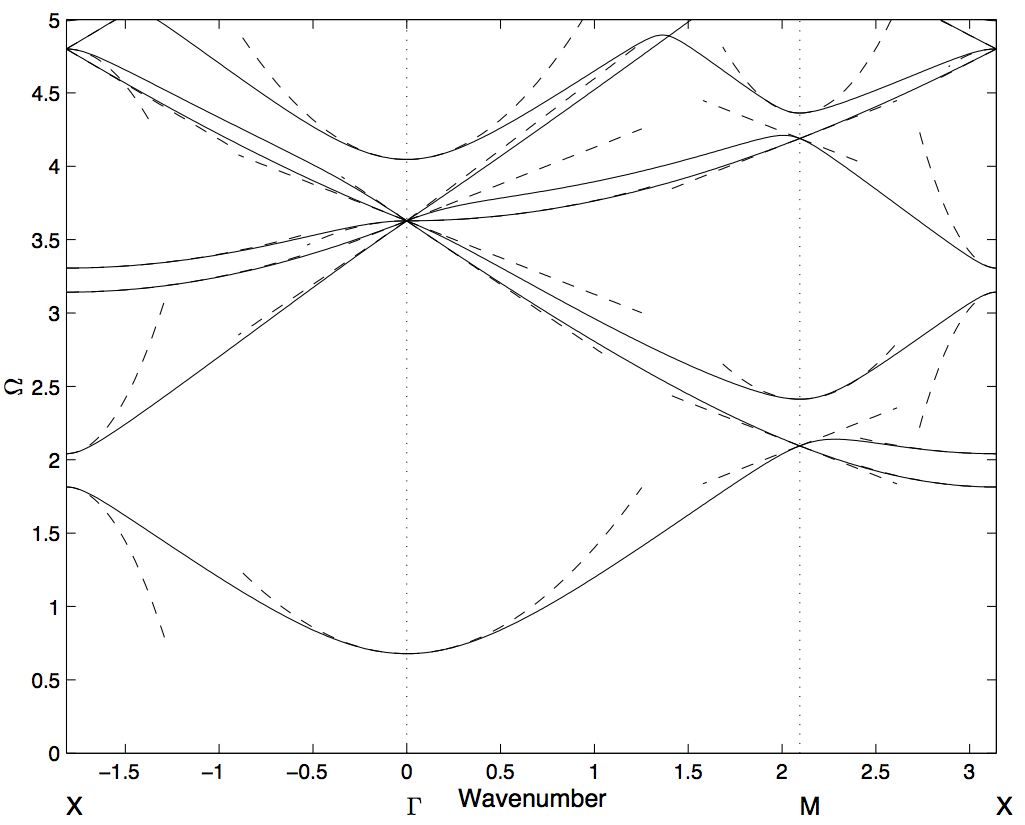}
\end{center}
\caption{\small {Band diagram for the hexagonal lattice of Dirichlet
    points in the Helmholtz case. Solid lines are from equation
    \eqref{eq:constrained_pt_disp_eqn}, with
    the HFH asymptotics shown as dashed curves.}}
\label{fig:tri_dispersion_total2}
\end{figure}

For the hexagonal lattice, the principal cell has only one inclusion at $(0,0)$.
The dispersion diagram, shown in figure
\ref{fig:tri_dispersion_total2}, notably exhibits a a quintuple generalised Dirac point
(asymptotically four lines and one quadratic curve) at $\Gamma$, and HFH faithfully captures the group velocity of these lines as well as the curvature of the quadratic. 

\indent A recurrent feature in literature is that of star-shaped, highly
directional wave propagation at specific frequencies, which has emerged
in experiments and theory in optics \cite{chigrin03a,craster12b}, and
is perhaps most strikingly seen in mass-spring lattice systems 
\cite{slepyan08a,langley97b,craster12b}, as well as in frame structures
\cite{colquitt12a,colquitt15a}. HFH can be used to interpret these effects
 through the tensor coefficient $T_{ij}$, and
we demonstrate this for both the the Helmholtz and the Kirchhoff-Love equation.

 \begin{figure}[htb!]
\begin{center}
    \includegraphics[scale=0.27]{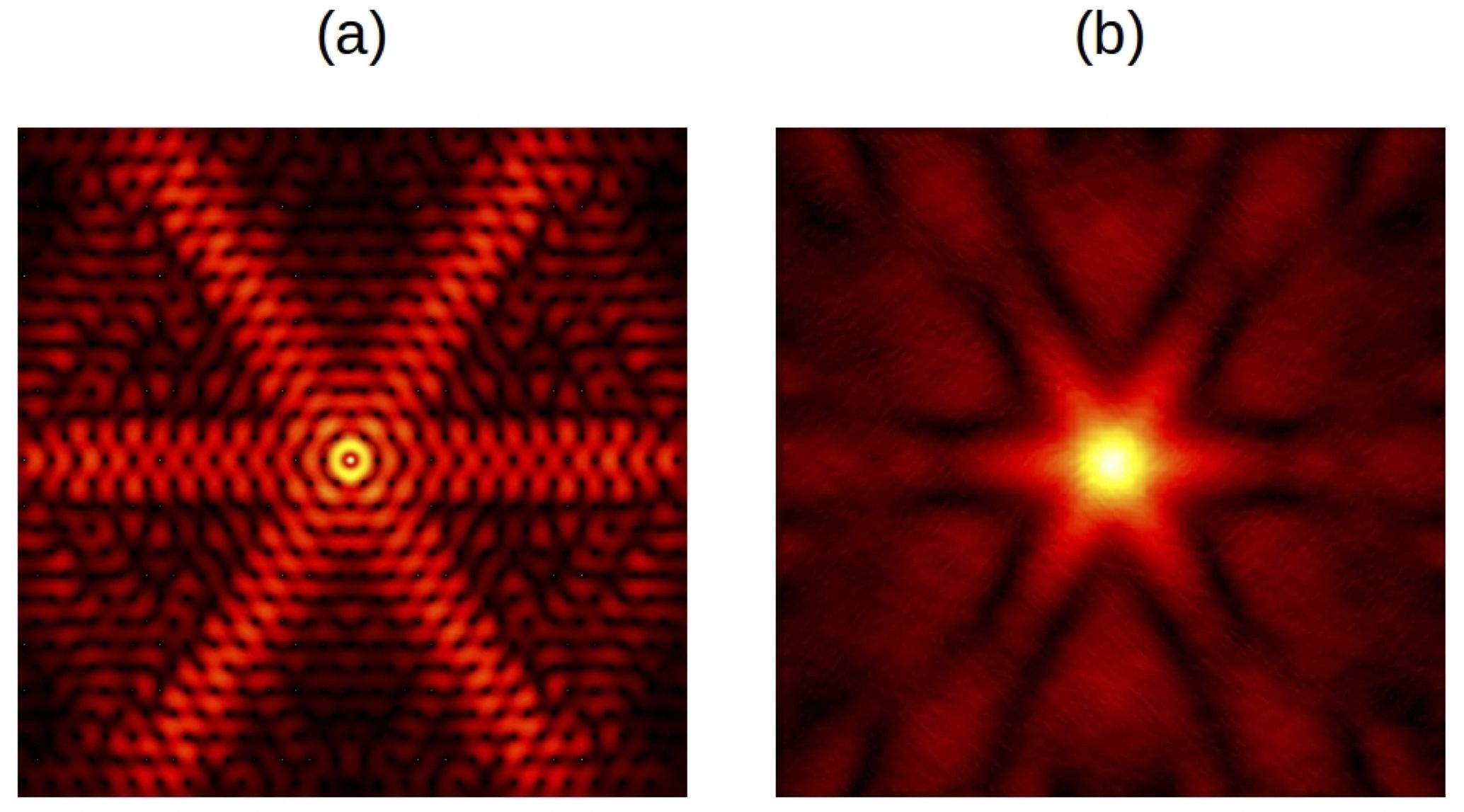}
\end{center}
\caption{\small {The Helmholtz case: A star shape is obtained by exciting the center of a hexagonal array of clamped points at frequency $\Omega=1.85$, near the first mode at point $X$ in fig. \ref{fig:tri_dispersion_total2}. Panel (a) is the full finite element simulation, where the source is approximated by a narrow Gaussian, and panel (b) is the HFH counterpart, also computed with finite elements, obtained by adding the field at point $X$ to its $\pm2\pi/3$ rotations. In both cases, PML is used to avoid reflections at the boundary of the domain.}}
\label{fig:star_1_85}
\end{figure}

 \begin{figure}[htb!]
\begin{center}
    \includegraphics[scale=0.27]{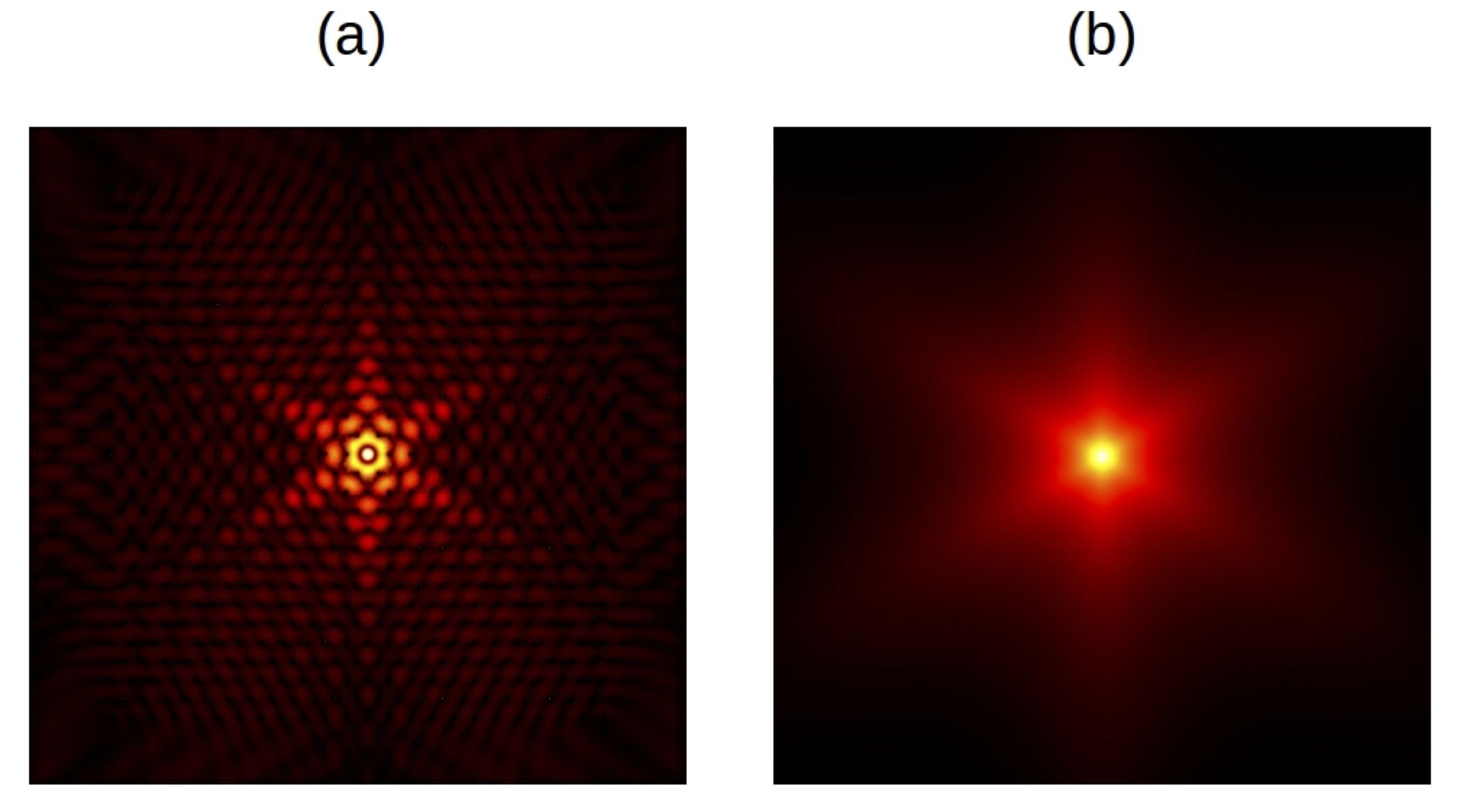}
\end{center}
\caption{\small {The Helmholtz case: A star shape is obtained by exciting the center of a hexagonal array of clamped points at frequency $\Omega=2.041$, near the second mode at point $X$ in fig. \ref{fig:tri_dispersion_total2}. Panel (a) is the full finite element simulation, where the source is approximated by a narrow Gaussian, and panel (b) is the HFH counterpart, also computed with finite elements, obtained by adding the field at point $X$ to its $\pm2\pi/3$ rotations. In both cases, PML is used to avoid reflections at the boundary of the domain.}}
\label{fig:star_2_041}
\end{figure}

\indent The dynamic anisotropy seen in figs. \ref{fig:star_1_85} and
\ref{fig:star_2_041} is qualitatively explained by the curvature of
the dispersion curves near point $X$ in
Fig. \ref{fig:tri_dispersion_total2}. Near the first band for
$\Omega\sim 1.85$ we have $T_{11}T_{22}<0$ (see table \ref{tab:hex_X}), signifying an effective PDE that is hyperbolic, not elliptic. The star shape is formed by waves that
are directed along the characteristics of this PDE. The angle between the characteristics is twice the inverse tangent of the ratio $\sqrt{|T_{11}/T_{22}|}$. The effect in fig. \ref{fig:star_2_041} appears similar but is fundamentally different. Both $T_{ii}$ coefficients are positive but $T_{22}/T_{11} \gg 1$. The propagation is thus directed along the $\mathbf{j}$-direction, and the sum of this with its $2\pi/3$ symmetry rotations yield the effect. This is identical to the effect seen for the
analogous discrete system \cite{makwana14}.

\begin{table}[t!]
\centering
\begin{tabular}{ccc}\hline
$T_{11}$ & $T_{22}$ & $ \Omega_0$ \\
$1$ & $-13.6253$ & $ 1.8141$\\
$0.8262$ &  $15.5661$ & $ 2.0409$\\
\hline
\end{tabular}
\caption{\small {The first two standing wave frequencies for a hexagonal cell with clamped holes at wavenumber ${\boldsymbol \kappa}=(0,\pi/\sqrt{3})$ at $X$,
 The above coefficients are used in Fig. \ref{fig:tri_dispersion_total2}}}
\label{tab:hex_X}
\end{table}

\subsubsection{Honeycomb array}

\begin{figure}[Hb!]
\begin{center}
    \includegraphics[scale=0.35]{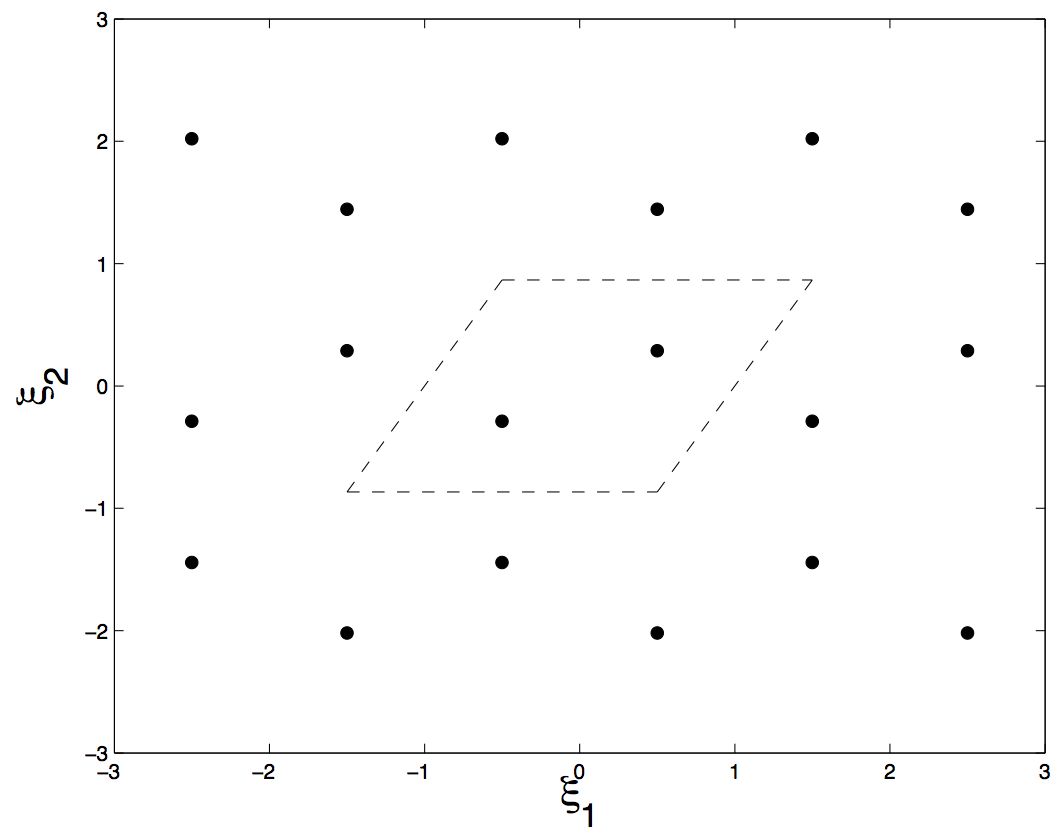}
\end{center}
\caption{\small {The honeycomb arrangement of inclusions with the
    elementary cell, containing two inclusions, shown by the dashed lines.}}
\label{fig:honey_arrangement}
\end{figure}

The honeycomb array can be obtained from the hexagonal lattice by including two inclusions in each unit cell (fig. \ref{fig:honey_arrangement}), located at points given by
\beq
{\bf I}^{(1,2)}=\pm\frac{1}{2\sqrt{3}}\left(\sqrt{3}{\bf i}+{\bf j} \right), 
\label{eq:hon_inclusion_posn} \eeq where $\pm$ correspond to the first and second inclusions respectively. Using equation \eqref{eq:constrained_pt_disp_eqn} we find the dispersion relation from the zeros of the determinant,
\beq
\varrho_{1,1} \pm |\varrho_{1,2}|=0,
\label{eq:hon_pt_disp} \eeq  where the reciprocal lattice vector is
\beq
{\bf G}=\pi \left[n, \frac{1}{\sqrt{3}}\left(2m-n \right) \right].
\label{eq:hon_recip_vec}\eeq

\begin{figure}[htb!]
\begin{center}
    \includegraphics[scale=0.6]{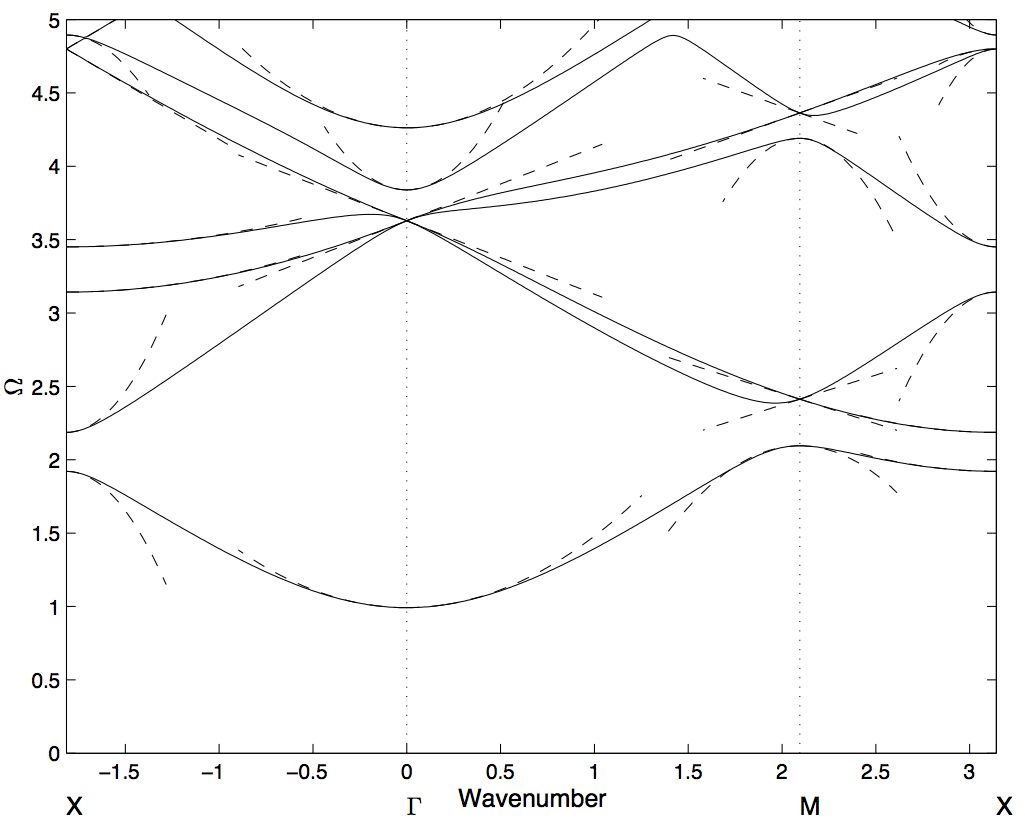}
\end{center}
\caption{\small {The band diagram for a honeycomb arrangement of
    Dirichlet inclusions in the Helmholtz case. Solid lines from
    equation \eqref{eq:hon_pt_disp} and the dashed lines are from the
    HFH asymptotics.}}
\label{fig:honey_dispersion_total2}
\end{figure}

\begin{figure}[htb!]
\begin{center}
    \includegraphics[scale=0.27]{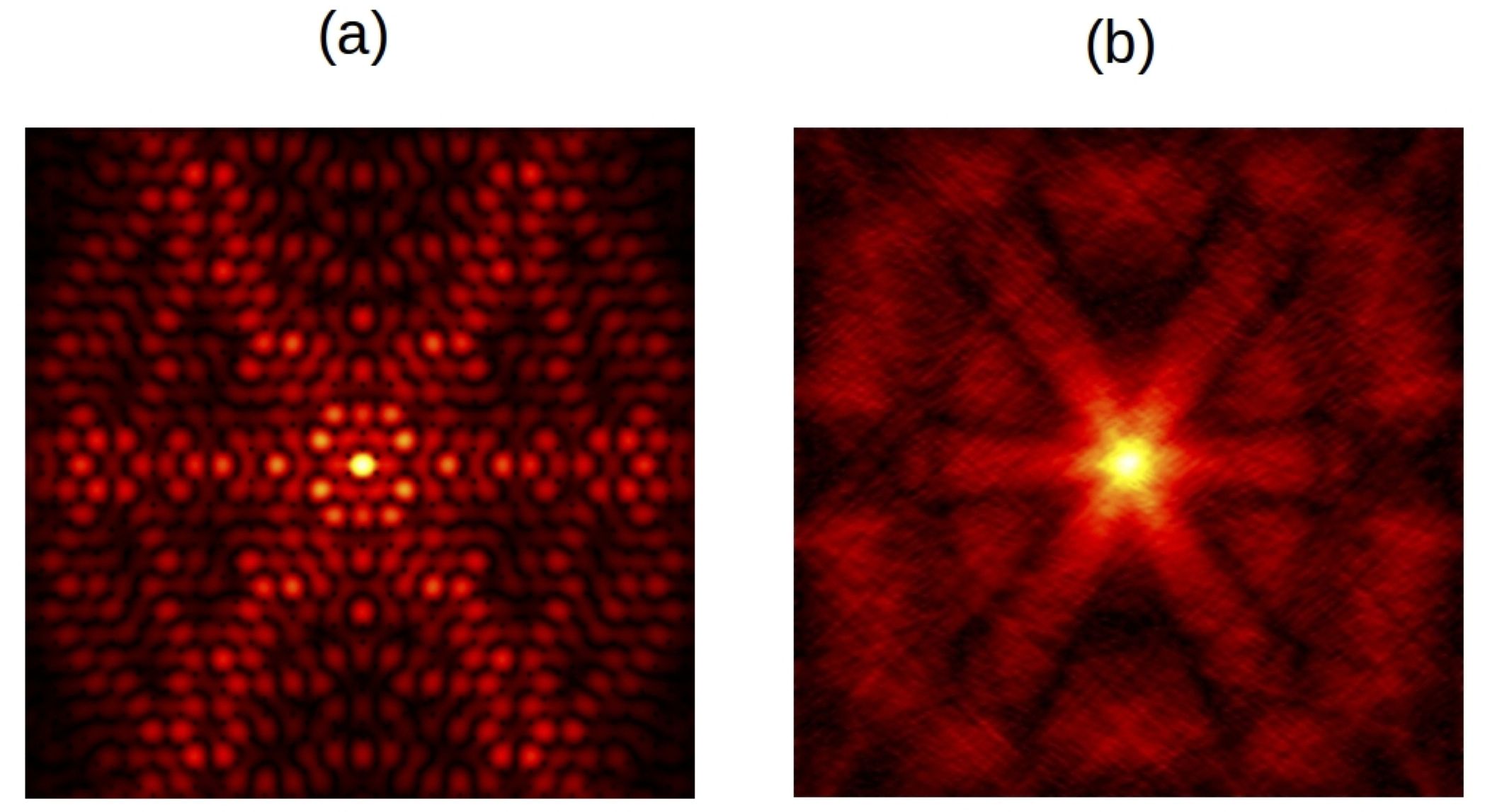}
\end{center}
\caption{\small {The Helmholtz case: A star shape is obtained by exciting the center of a honeycomb  array of Dirichlet points at frequency $\Omega=1.938$, near the first mode at point $X$ in fig. \ref{fig:honey_dispersion_total2}. Panel (a) is the full finite element simulation, where the source is approximated by a narrow Gaussian, and panel (b) is the HFH counterpart, also computed with finite elements, obtained by adding the field at point $X$ to its $\pm2\pi/3$ rotations. In both cases, PML is used to avoid reflections at the boundary of the domain.}}
\label{fig:star}
\end{figure}

\begin{figure}[h!]
  \centering
   \centerline{\includegraphics[scale=0.232]{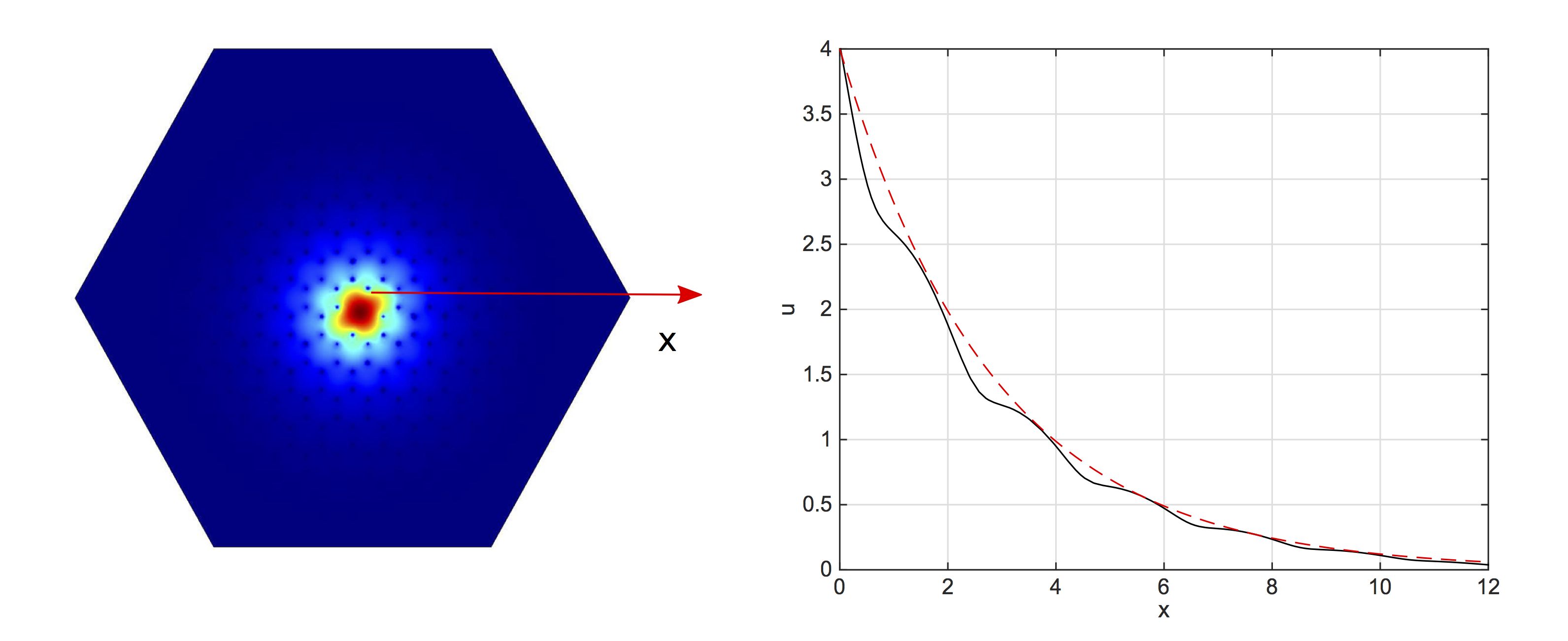}}
\caption{\small{The Helmholtz case: Localised defect mode at $\Omega=0.93$, induced by the removal of two adjacent pins, which lies in the zero-frequency stop-band just beneath the lowest eigenvalue at $\Gamma$. The effective equation is $0.97f_{0,xx}+0.97f_{0,yy}+(\Omega^2-\Omega_0^2)f_0=0$, where $\Omega_0=0.992$, which leads to an isotropically decaying envelope (the red dashed curve).}}
\label{fig:defect1}
\end{figure}

\begin{figure}[h!]
  \centering
   \centerline{\includegraphics[scale=0.232]{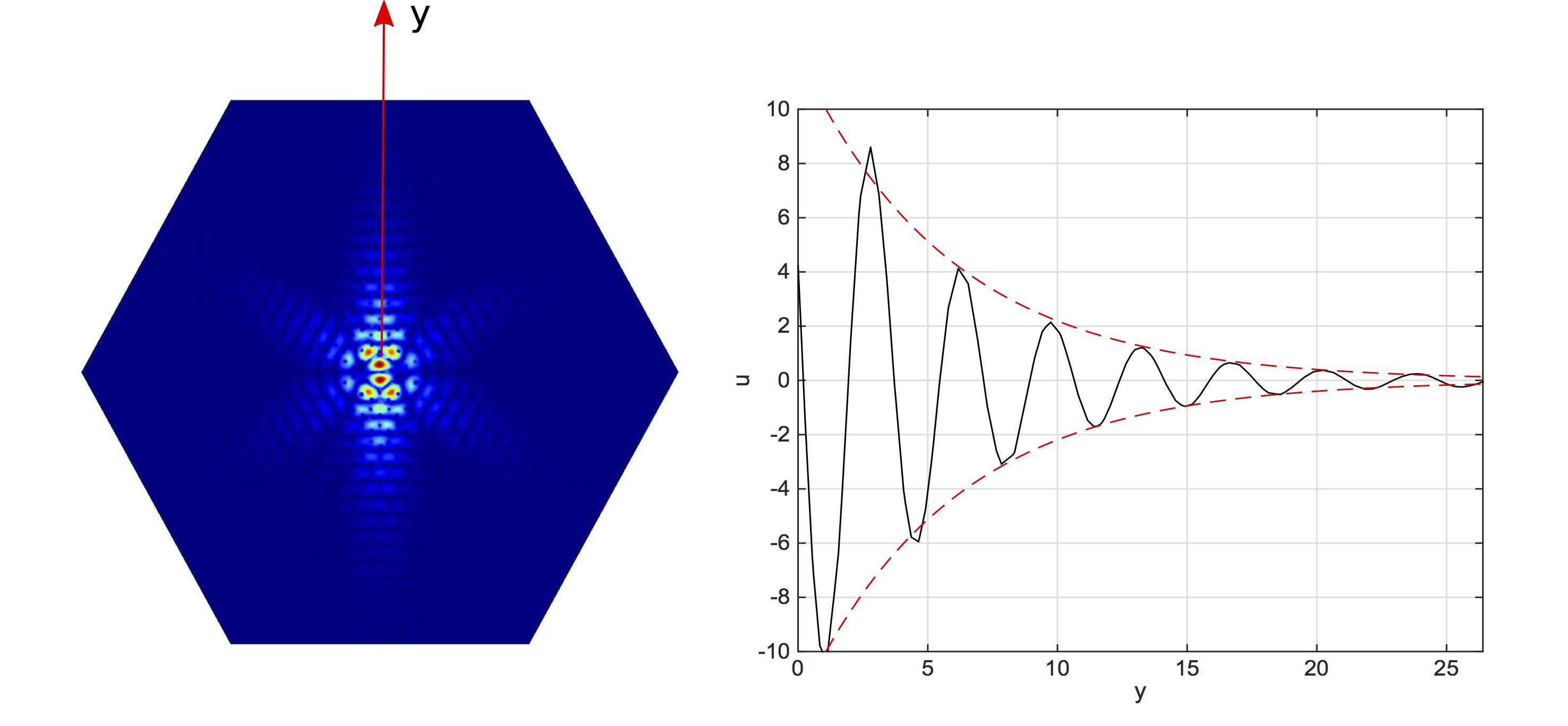}}
\caption{\small{The Helmholtz case: Localised defect mode at $\Omega=2.16$, induced by removing six adjacent pins, which lies in the narrow stop-band $2.09<\Omega<2.19$. The dominant effective equation is $0.95f_{0,xx}+12.41f_{0,yy}+(\Omega^2-\Omega_0^2)f_0=0$, where $\Omega_0=2.187$, corresponding to the second lowest eigenvalue at $X$, which leads to highly directed leakage in the vertical direction, and hence in 6 directions by rotational symmetry of the array. Note that this is one of two independent defect modes at this frequency, and the directivity can be primarily along any one of the 6 symmetric directions, corresponding to different linear combinations of these two.}}
\label{fig:defect2}
\end{figure}

The dispersion diagram for the honeycomb array, shown in Fig. \ref{fig:honey_dispersion_total2}, shares several features with that of the hexagonal array (Fig. \ref{fig:tri_dispersion_total2}). It too exhibits a generalised Dirac point at $\Gamma$, but in this case with one fewer branch, and also contains saddle points that yield hyperbolic behaviour and the characteristic star shapes. An additional feature here is the small omni-directional band gap for $2.09<\Omega<2.19$, which has implications for the existence of localised defect modes in the honeycomb structure. Localised defect states were analysed in detail
for the discrete analogue of HFH in \cite{makwana13}, and here we consider the effect of introducing a finite defect by means of removing one or more pins from
the honeycomb array. If the size and shape of the defect is
appropriately chosen, and the perturbed array treated in the context
of an eigenvalue problem, we observe a localised state, in
which the field decays evanescently in the surrounding medium. The
decay rate of the envelope function is then governed by equation (\ref{eq:TijEquation}).  Our effective medium approach may be applied inside
any stop-band, and here we demonstrate this both for the
zero-frequency gap (fig. \ref{fig:defect1}) and also inside the narrow gap $2.09<\Omega<2.19$ (fig. \ref{fig:defect2}). There is currently much interest
in such localised modes, for example in
the context of opto-mechanical problems  \cite{gavartin11}, where the simultaneous
localisation of electromagnetic and elastic modes has applications
for, among other things, optical cooling. Analysis of localised defect
states is also fundamental to the field of photonic crystal fibres
\cite{zolla05a}. 

\subsubsection{Rhombic lattice}
\begin{figure}[htb!]
\begin{center}
    \includegraphics[scale=0.75]{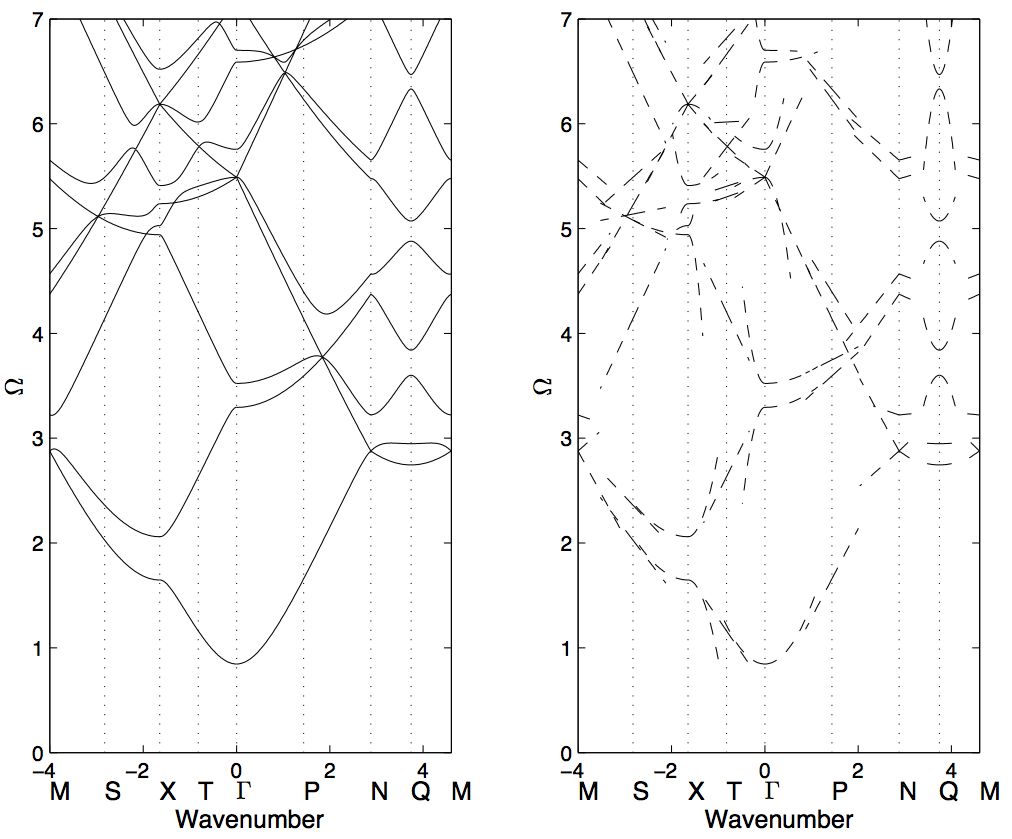}
\end{center}
\caption{The band diagram for the rhombic lattice in the Helmholtz
  case: Left panel showing with solid lines the exact dispersion
  relation for the rhombic lattice ($\alpha=0.7$). Asymptotics are
  shown in the right panel as dashed lines and found by perturbing away from the corners of the Brillouin zone, $\Gamma$, $N$, $M$ and $X$ as well as the mid-points of $\Gamma N$, $NM$, $MX$, $M\Gamma$, namely $P$, $Q$, $S$, $T$ respectively.}
\label{fig:rhombic_disp}
\end{figure}

\begin{figure}[h!]
\begin{center}
    \includegraphics[scale=0.6]{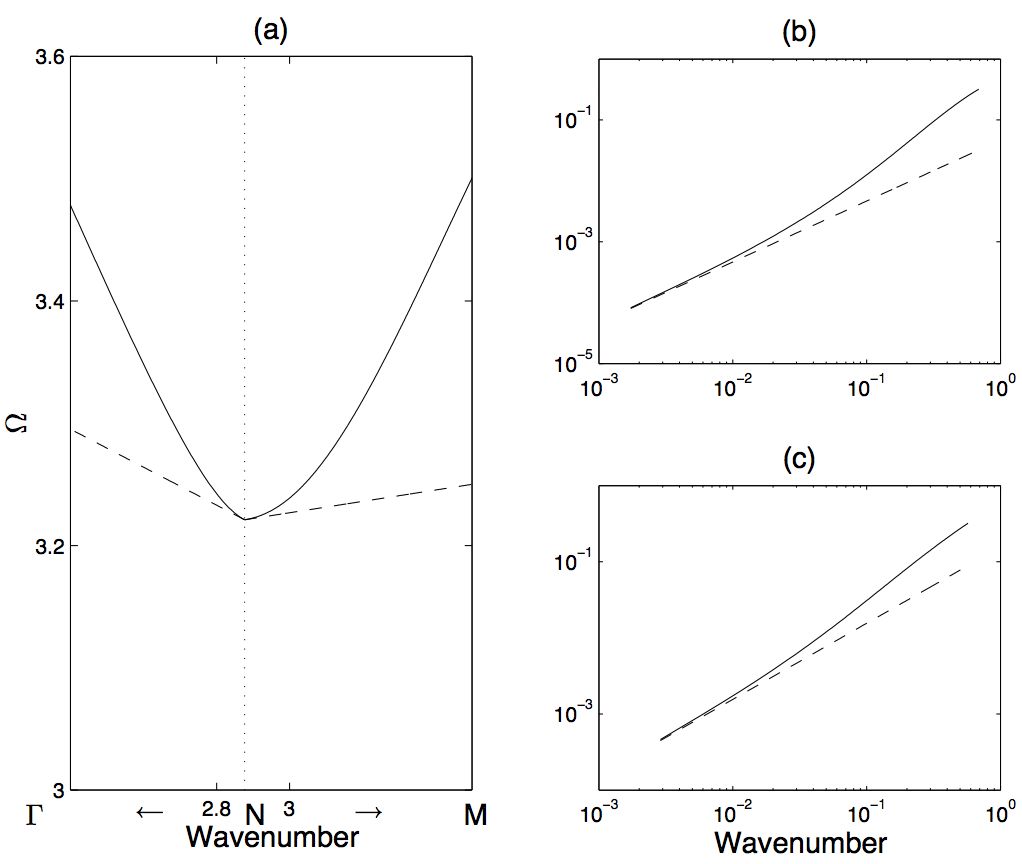}
\end{center}
\caption{Close-up plot of third branch of the band diagram of
  fig. \ref{fig:rhombic_disp}  with asymptotics (dashed) as well as loglog plots. Panel (a) shows the close-up of the third branch near point $N$ of the Brillouin zone. In panel (a) the power law of the dispersion curve is unclear but panels (b) and (c) show that the latter is linear and the asymptotics by HFH match well the dispersion curves.}
\label{fig:loglog_N}
\end{figure}

\begin{figure}[h!]
\begin{center}
    \includegraphics[scale=0.6]{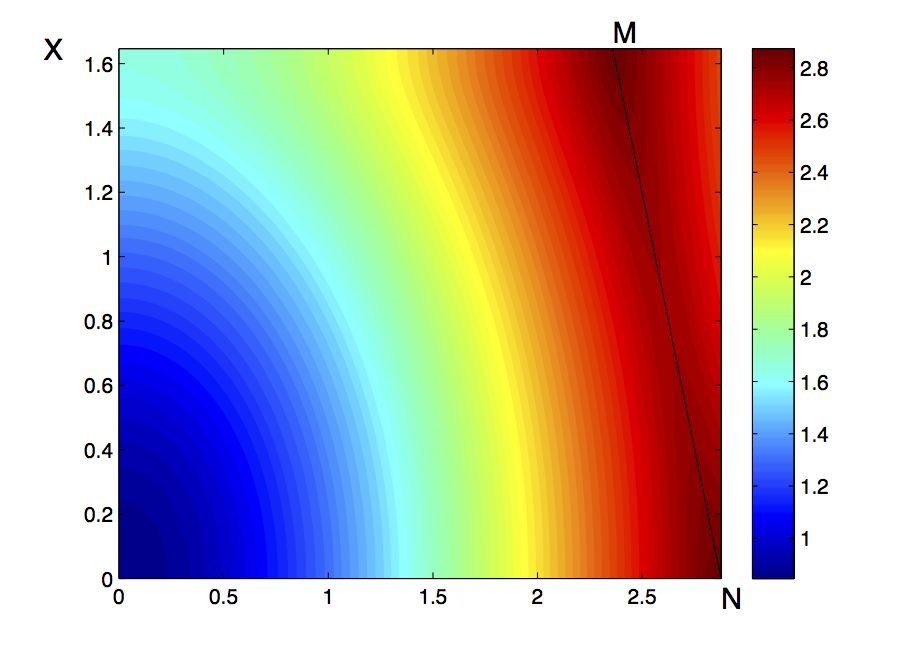}
\end{center}
\caption{Isofrequency curve for first mode of rhombus geometry which demonstrates symmetry along the NM path}
\label{fig:rhombic_isofreq}
\end{figure}

The final lattice we consider is the least symmetric of our examples, and the resulting irreducible Brillouin zone has four vertices, rather that three. Fig. \ref{fig:rhombic_disp} shows the dispersion diagram for the rhombic lattice, and demonstrates how HFH captures the group velocity and curvature at any point in the Brillouin zone. As the discretisation of the Brillouin zone becomes more dense, the exact dispersion curves can be retrieved via the asymptotic method. As an aside, an interesting nuance is observed whereby the dispersion curves are symmetric about $Q$ along the path $NM$. Further investigation reveals that this symmetry is only present along that particular path, as is clear from the isofrequency plot Fig. \ref{fig:rhombic_isofreq}. The dispersion curves near points $N$ and $M$ are linear as it is shown by Fig. \ref{fig:loglog_N}, despite appearing to have locally quadratic behaviour. The group velocity is non-zero at these points, even though the dispersion surface admits local extrema.   

\newpage
\subsection{Kirchhoff-Love equation}
\subsubsection{Hexagonal lattice}

\begin{figure}[h!]
\begin{center}
    \includegraphics[scale=0.76]{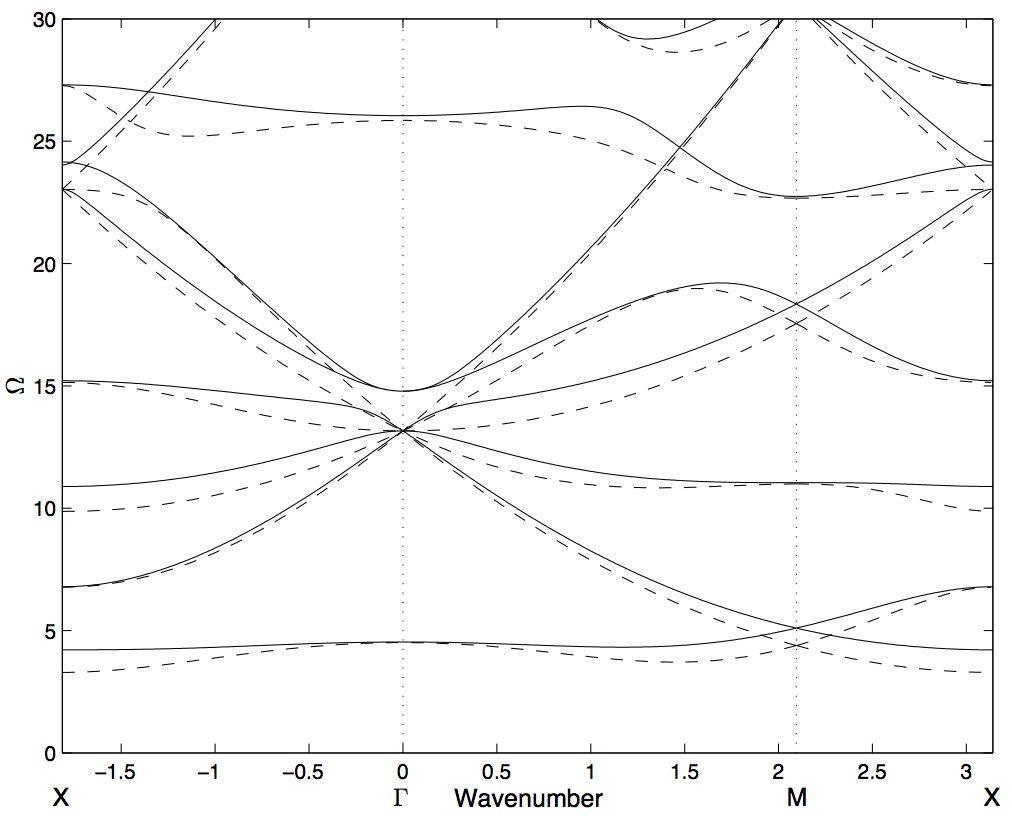}
\end{center}
\caption{\small {Dispersion curves for the Kirchhoff-Love model with inclusions arranged in a hexagonal lattice. Solid curves are for finite radius inclusions ($R=0.01$) whilst the dashed lines are for the pinned points, equation \eqref{eq:constrained_pt_disp_eqnK}.}}
\label{fig:hex_KL_total}
\end{figure}

In the case of the Kirchhoff-Love model, we opt to solely examine inclusions arranged in a hexagonal lattice. Similar features dispersive features are observed, An interesting distinction between the curves for the Helmholtz equation and those of Kirchhoff-Love, is that the latter model is far more sensitive to changes in the radii of the inclusions. This can be seen in the dispersion curves (fig. \ref{fig:hex_KL_total}), where the solid curves are for finite (but very small) radius inclusions ($R=0.01$) and the dashed curves are for zero-radius holes, found using the exact solution, equation \eqref{eq:constrained_pt_disp_eqn}. This nuance can be explained by comparing the two dispersion curves, (figures \ref{fig:tri_dispersion_total2} and \ref{fig:hex_KL_total}) and noting that the frequencies in the plate model are considerably higher than those found using Helmholtz equation. Hence as the frequency increases the dashed lines deviate further from the solid lines.

\begin{table}[h!]
\centering
\begin{tabular}{cccc}\hline
$T_{11}$ & $T_{22}$ & $ \Omega_0$ & \text{Radius} \\
$37.9829$ & $-35.8258$ & $ 6.7903$ & $ 0.01$\\
$71.3788$ &  $-29.1596$ & $ 6.7733$ & \text{Pinned}\\
\hline
\end{tabular}
\caption{ The $T_{ij}$ coefficients found in equation \eqref{eq:tijK} and frequency values for both the finite radius holes and for the pinned points are shown for the second mode at point X in the dispersion curve (fig. \ref{fig:hex_KL_total}). The values for the pinned points are taken from equation \eqref{eq:gen_TijK}. Note the strong mismatch between the $T_{ij}$ coefficients for small clamped inclusions and pinned points.}
\label{tab:hex_KL}
\end{table}

\begin{figure}[Htb!]
\begin{center}
    \includegraphics[scale=0.3]{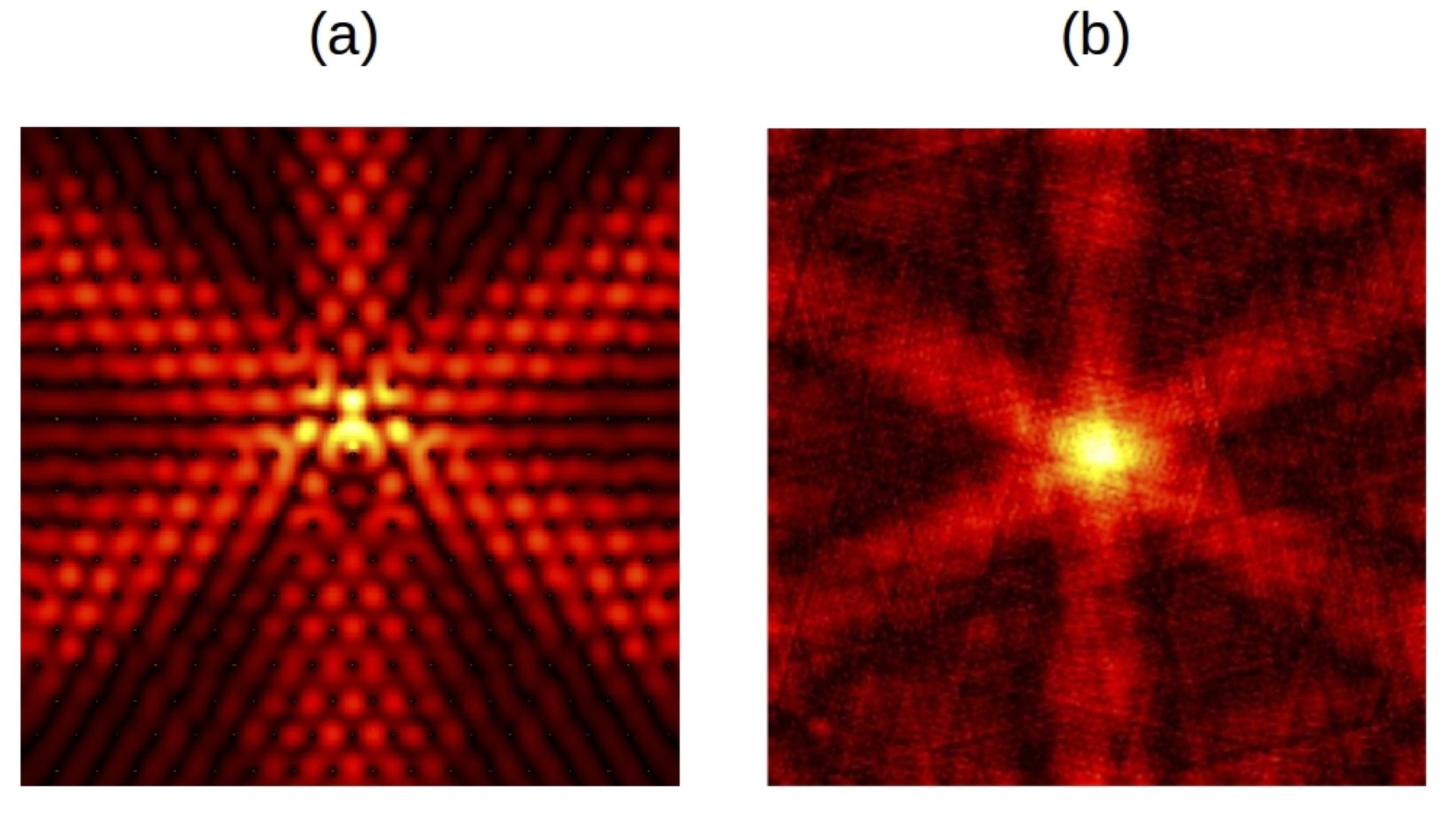}
\end{center}
\caption{A star shape is obtained by exciting the center of a hexagon array of clamped points at frequency $\Omega= 6.7903$ for the Kirchhoff-Love model, near the second mode at point X in figure \ref{fig:hex_KL_total}. Panel (a) shows the plate simulation, whilst panel (b) is the HFH effective medium simulation, both are for the finite radius of $R=0.01$. Similar to the Helmholtz case the three characteristics are obtained due to the inherent three-fold symmetry of the structure.}
\label{fig:plates_star}
\end{figure}

\indent The strong discrepancies between a small change in radius is demonstrated in table \ref{tab:hex_KL} for the second mode at point X in fig. \ref{fig:hex_KL_total}. This mode is of interest, as excitation about this frequency yields star shape oscillations (fig. \ref{fig:plates_star}) similar to those seen earlier for the Helmholtz equation. This oscillatory pattern can be equivalently ascertained by taking into account the inherent three-fold symmetry of our medium and the hyperbolic PDE obtained using HFH.

\section{Concluding remarks}
\label{sec:conclusions}
It is clear that a microstructured medium with any periodic arrangement of inclusions can now be homogenised for any frequency in the Bloch spectrum  and effective continuum equations deduced. In contrast to the previously studied case of orthogonal lattices \cite{antonakakis13b}, a key technical difficulty is that the microscale and macroscale are naturally in different coordinate systems. Once this issue has been overcome the effective equations are versatile and capture, for instance, the strongly directional anisotropic behaviour at critical frequencies. The validity and usefulness of our homogenisation method is demonstrated with the topical honeycomb arrangement of inclusions, whereby we introduce defects on the microscale and show that the envelope modulation in the surrounding medium is perfectly captured by our long-scale effective PDE. Additionally, due to the mathematical similarity between Helmholtz and Kirchhoff-Love equations, we apply our analysis to both of these models, and find that the former helps us to better understand the latter. Finally, exact solutions are constructed for constrained points using Fourier series and these are used to provide analytical expressions within our asymptotic framework.
\newpage
\bibliographystyle{siam}
\bibliography{refe}

\end{document}